\documentclass[%
 reprint,
 amsmath,amssymb,
 aps,
prr,
]{revtex4-2}
\usepackage{braket,color,graphicx,float}
\usepackage{changes,mathtools}
\DeclareUnicodeCharacter{2212}{\textendash}

\begin{document}
\title{Revealing Ultrafast Phonon Mediated Inter-Valley Scattering through Transient Absorption and High Harmonic Spectroscopies}

\preprint{APS/123-QED}
\def\doubleunderline#1{\underline{\underline{#1}}}


\author{Kevin Lively}
\affiliation{Max Planck Institute for the Structure and Dynamics of Matter, Luruper Chaussee 149, 22761 Hamburg, Germany}
\affiliation{German Aerospace Center (DLR), Institute for Software Technology, Department of High-Performance Computing, Rathausallee 12, 53757 Sankt Augustin, Germany}
\author{Shunsuke A. Sato}
\affiliation{Center for Computational Sciences, University of Tsukuba, Tsukuba, Ibaraki 305-8577, Japan}
\affiliation{Max Planck Institute for the Structure and Dynamics of Matter, Luruper Chaussee 149, 22761 Hamburg, Germany}
\author{Guillermo Albareda}
\affiliation{Max Planck Institute for the Structure and Dynamics of Matter, Luruper Chaussee 149, 22761 Hamburg, Germany}
\affiliation{Ideaded, Carrer de la Tecnologia, 35, 08840 Viladecans, Barcelona, Spain}
\author{Angel Rubio}
\affiliation{Max Planck Institute for the Structure and Dynamics of Matter, Luruper Chaussee 149, 22761 Hamburg, Germany}
\affiliation{Center for Computational Quantum Physics (CCQ), Flatiron Institute, New York 10010, New York, USA}
\author{Aaron Kelly}
\affiliation{Max Planck Institute for the Structure and Dynamics of Matter, Luruper Chaussee 149, 22761 Hamburg, Germany}
\affiliation{Hamburg Center for Ultrafast Imaging, Universit\"at Hamburg, Luruper Chaussee 149, 22761 Hamburg, Germany}
\email[Author to whom correspondence should be addressed: ]{aaron.kelly@mpsd.mpg.de}
\date{\today}

\begin{abstract}
Processes involving ultrafast laser driven electron-phonon dynamics play a fundamental role in the response of quantum systems in a growing number of situations of interest, as evidenced by phenomena such as strongly driven phase transitions and light driven engineering of material properties. To show how these processes can be captured from a computational perspective, we simulate the transient absorption spectra and high harmonic generation signals associated with valley selective excitation and intra-band charge carrier relaxation in monolayer hexagonal boron nitride. We show that the multi-trajectory Ehrenfest dynamics approach, implemented in combination with real-time time-dependent density functional theory and tight-binding models, offers a simple, accurate and efficient method to study ultrafast electron-phonon coupled phenomena in solids under diverse pump-probe regimes which can be easily incorporated into the majority of real-time \textit{ab initio} software packages.
\end{abstract}

\maketitle

\section{Introduction}

Time resolved spectroscopies, such as time and angle resolved photoemission, time resolved photoluminesence and transient absorption spectroscopy (TAS) constitute fundamental tools to study the flow of energy in materials following excitation by light. Understanding the microscopic details of the excitation and relaxation pathways can serve as the basis for deterministic manipulation of material properties for technological applications such as enhanced photodetectors \cite{Tielrooij2015, Trovatello2022}, long lived optically controlled qubit registers \cite{Mak2012,Liu2019} and attosecond control of magnetic ordering for ultrafast spintronics \cite{Dewhurst2018,Siegrist2019,Neufeld2023}. In parallel to developments pushing the time, energy, and momentum resolution of these spectroscopic techniques, there has been a plethora of phenomena studied under novel conditions such as transient phases and Floquet renormalization under strong parametric driving \cite{Mitrano2016,Huebner2018,Nova2019,Li2019,Disa2023}, chemical reaction rate modification under exposure to cavity confined fields \cite{Frisk2019,Fregoni2022}, and exotic quantum phases when layering 2D materials \cite{Li2021,Wu2019}. The study of the microscopic origins of these phenomena pushes the boundaries of theoretical tools which are useful near equilibrium conditions, in particular for one of the most fundamental processes for understanding the behavior of materials: the electron-phonon interaction. 

Treating coupled electron-phonon dynamics in simulations of periodic systems is typically limited to either phenomenological coupling and decay terms, or coarse approximations such as the two-temperature model \cite{Sato2019,Mao2022,Caruso2022}. Going beyond these options leads one to consider an explicit treatment of the phonon degrees of freedom, which can be achieved through the time-dependent Boltzmann equation (TDBE) \cite{Malic2011,Bernardi2016,Zhou2021,Caruso2021,Tong2021,Caruso2022}, which is derived in the perturbative limit of the electron-phonon interaction. Attempts to move beyond some of the constraints of the TDBE include approaches based on density matrix formalism, such as the Bloch equation \cite{Winzer2012,Winzer2013,Berghauser2018}, the Hierarchical Equations of Motion \cite{Jankovic2022} or the time-dependent Density Matrix Renormalization Group \cite{Li2021_2,Sous2021} for 1D systems, and non-equilibrium Green's function approaches based on solving the Kadanoff-Baym equations \cite{Molina-Sanchez2017,Schlunzen2020,perfetto2023}. 

Conversely, the electronic system can be treated in a fully \textit{ab initio} manner and one can still capture the effect of phonon fluctuations, even in cases with very strong coupling, via static displacement approaches which sample phonon distortions in large supercells \cite{Zacharias2016,Zacharias2020}. In the adiabatic limit, this approach has recently been shown to be equivalent to the Feynman expansion to all orders of the electron-phonon interaction perturbation \cite{Nery2022}, and is analogous to the nuclear ensemble average technique from molecular physics, where electronic properties are obtained by averaging over the nuclear coordinate distribution on the ground Born-Oppenheimer state \cite{Crespo-Otero2012}. 
Breaking the high symmetry equilibrium lattice structure of periodic systems has been found to significantly alter the results of ab-initio calculations; disorder has been argued as being responsible for capturing a significant portion of what is usually attributed to electron correlation when using ostensibly uncorrelated DFT methods \cite{Zunger2022}, as well as being the dominant factor in phase transitions typically argued to be rooted in electronic structure \cite{Baldini2023}.

A natural step beyond including the static disorder of the nuclei is to also consider their dynamics, which opens the possibility to treat time-dependent response properties where the nuclear forces are dependent on nonequilibrium electronic configurations. A mean field treatment that can be applied in this context is the multi-trajectory Ehrenfest approach (MTEF), which allows one to recover the quantum statistics of the equilibrium nuclear subsystem \cite{Grunwald2009} whilst approximating the time-evolution using Ehrenfest trajectories. This approach has been shown to capture Franck-Condon physics \cite{Lively2021} as well as time-resolved out of equilibrium system dynamics \cite{Hoffmann2019,Krumland2022,Brink2022}. Propagating the system in real time allows for coherent electronic evolution at short time scales, while accounting for all orders of interaction with both external fields and the phonon system (at the mean field level). 

Although the basic ingredients required for MTEF are already available in most real-time \textit{ab initio} codes, and despite some existing formulations of reciprocal space semi-classical dynamics in the literature \cite{Li2017, Sato2018, Krotz2021}, to the authors' best knowledge, application of this method has not been widely explored in periodic systems, nor has the static displacement approach been widely applied to time-resolved phenomena. Rather, most uses of \textit{ab initio} semi-classical dynamics in periodic systems involve only a single trajectory \cite{Shinohara2010,Hu2022,Neufeld2022}, typically using a primitive unit cell or initializing the phonon distortion with classical molecular dynamics simulations which often fail to capture the exact quantum statistics of the initial state \cite{Freeman2022}. Nonetheless, inclusion of more phonon modes through supercell dynamics allows for detailed study of fundamental processes such as the relaxation of excited electrons through phonon emission, anharmonic phonon-phonon scattering and phase transitions, even at the single trajectory level \cite{Guan2022,Guan2022_2}. 

Therefore, in this work we formulate the Wigner representation for the phonon subsystem in a generic manner using \textit{ab initio} dispersion relations of real materials, yielding a framework that systematically captures the equilibrium properties of the phonon system. While MTEF is well known to suffer from zero point energy (ZPE) leakage and incorrect thermalization between the quantum and classical systems at long time scales, there are several schemes that can systematically correct these failures with added computational cost \cite{Kelly2015,Hsieh2023}. However, as our focus here is limited to studies of the short time-scale dynamics of strongly driven systems, we will not pursue such corrections and simply aim to discover what can be achieved at the mean field level.  

As an illustrative example we study ultrafast phonon mediated electronic reorganization following valley selective laser excitation in monolayer hexagonal boron nitride (hBN), demonstrating the ability of MTEF to capture many of the essential features of the process which has been argued to be a significant driver of valley selective relaxation in Transition Metal Dichalcogenides (TMDs) \cite{Zeng2012,Selig2016,Molina-Sanchez2017,Xu2021}. Due to negligible spin-orbit coupling in hBN, the two $K$ valleys in the Brillouin Zone (BZ) are degenerate, while valley specific selection rules are preserved upon interaction with circularly polarized light \cite{Geondzhian2022,Hewageegana2021,Jimenez-Galan21, Jimenez-Galan20}, making this a prototypical material to study the relaxation of excited charge carriers theoretically. Using a real space \textit{ab initio} supercell approach with time-dependent density functional theory (TDDFT), and a reciprocal space tight binding (TB) model, we find that including phonon fluctuations leads to a rapid redistribution of excited charge carriers within a characteristic time scale of less than 30 fs, and that these results converge with as little as two trajectories.

We show that in the static limit, the harmonic Wigner distribution of phonon momenta and coordinates reduces to the phonon distribution obtained in the reciprocal space picture of Williams-Lax theory \cite{Zacharias2020}. We compare the results of propagating the electronic system with MTEF to the limit of frozen phonons, as well as the TDBE approach, finding broad qualitative agreement between these approaches across temperatures from $0-2000$ K while reproducing the experimentally observed low-temperature behaviour of analogous TMD systems. Finally, we demonstrate the flexibility of our method to predict and recreate spectroscopic experiments by simulating two different transient absorption measurements. First we propose a circularly polarized TAS measurement and demonstrate the signal corresponds directly to the valley asymmetry decay. Second, we replicate a recently performed study \cite{Mitra2023} using the ellipticity of harmonics generated under extreme laser driving to track tunable valley selective excitation in hBN using bichromatic counter-rotating `trefoil' pumps, also showing a rapid decay of the observable signal. Both cases demonstrate the ease with which our approach can be applied within tight binding and fully \textit{ab initio} real time electronic dynamics software packages under arbitrary pump-probe setups.

\section{Theory}
Here we briefly summarize how the description of the phonon subsystem can be formulated based on the phonon dispersion of real materials within MTEF by appealing to the Wigner distribution of the phonon system. We then elaborate on the connections between this approach and the static displacement formalism, and we briefly outline the TDBE that will be used for later comparison. We generally use atomic units throughout, though sometimes for clarity $\hbar$ is written explictly. 

\subsection{Multitrajectory Ehrenfest dynamics}
While a variety of routes to derive the Ehrenfest equations of motion are available, we focus here on how the MTEF approach can be derived as the uncorrelated solution to the quantum-classical Liouville equation (QCLE) \cite{Aleksandrov1981,Kapral1999}, which describes the approximate time-evolution of the total density matrix. The partial Wigner transform is employed, which for an arbitrary operator is defined as:
\begin{equation}\label{eq:Wigner Transform}
    \hat{O}_{W}(\mathbf{R,P}) = (2\pi)^{-dN}\int d\mathbf{Q} e^{-i\mathbf{P}\cdot\mathbf{Q}}\braket{\mathbf{R}+\mathbf{Q}/2|\hat{O}|\mathbf{R}-\mathbf{Q}/2}.
\end{equation}
Here the list of variables $(\mathbf{R},\mathbf{P})$ represents the full set of nuclear position $\mathbf{R} = \left(\mathbf{R}_1,\ldots,\mathbf{R}_N\right)$ and momentum $\mathbf{P}= \left(\mathbf{P}_1,\ldots,\mathbf{P}_N\right)$ variables which are vectors in $N\times d$ Cartesian dimensions, and we note that the operator character of objects in the electronic Hilbert space is unchanged by the partial Wigner transform.  

In the mean field limit the density operator can be factorized $\hat{\rho}_W = \rho_{\text{n},W}(\mathbf{R,P},t)\hat{\rho}_{\text{e}}(t)$, and the Wigner function of the nuclear degrees of freedom can be represented by an ensemble of $N_t$ independent trajectories, $\rho_{\text{n},W}(\mathbf{R,P},t) = \sum_{i=1}^{N_t} w_i \delta(\mathbf{R} -\mathbf{R}_i(t))\delta(\mathbf{P} -\mathbf{P}_i(t))$, with weights $w_i$. The time evolution of the electronic density and the phase space coordinates is given by the Ehrenfest equations of motion \cite{Grunwald2009}:
\begin{equation}\label{eq:MTEF eom}
    \begin{split}
    \partial_t\hat{\rho}_{e}(t) &= -i\left[\hat{H}_W(\mathbf{R}_i(t),\mathbf{P}_i(t)),\hat{\rho}_{e}(t)\right]\\
    \dot{\mathbf{P}}_i &= -\text{Tr}\left[\hat{\rho}_{e}(t)\mathbf{\nabla}_{\mathbf{R}}\hat{H}_W(\mathbf{R,P})\bigg|_{(\mathbf{R}_i(t),\mathbf{P}_i(t))}\right]\\
    \dot{\mathbf{R}}_i &= \frac{\mathbf{P}_i}{\mathbf{M}},
    \end{split}
\end{equation}
where $\mathbf{M}$ are the nuclear masses. These equations are solved for the independent trajectories with initial conditions $(\mathbf{R}_i(0),\mathbf{P}_i(0))$ sampled from $\rho_{\text{n},W}$. Observables are constructed by averaging over the ensemble of trajectories; in the case of equal trajectory weights, $\braket{O(t)} = \frac{1}{N_t} \sum_{i=1}^{N_t} \text{Tr}\left[\hat{O}_W(\mathbf{R}_i(t),\mathbf{P}_i(t))\hat{\rho}_W(\mathbf{R}_i(0),\mathbf{P}_i(0))\right]$. 

\subsection{The Phonon Subsystem}
For completeness we restate some textbook definitions of the phonon coordinates, in particular drawing from the work of Br\"{u}esch \cite{Bruesch1982} and Giustino \cite{Giustino2017}, with special emphasis on the often less well described conjugate momenta, which play an important role in the MTEF method (however, for a notable exception see \cite{Caruso_Zacharias2022}).

For a supercell composed of $N_p=N_1\times\ldots\times N_l$ primitive cells in $l$ periodic dimensions, we utilize the Born von-Karman (BvK) boundary conditions (see \ref{section: SI BvK}). Each primitive cell contains $N_c$ unique atoms which for a given lattice configuration have equilibrium positions in the primitive cell of $\mathbf{R}_{\alpha}^0$, for $\alpha=1,\ldots, N_c$. The equilibrium position of a given atom $\alpha$ within the supercell is specified by the primitive cell position $\mathbf{R}_p$; for primitve cell index $p = 1,\ldots, N_p$, and the primitive cell equilibrium position, $\mathbf{R}_{\alpha p}^0 = \mathbf{R}_p + \mathbf{R}_{\alpha}^0$. We further denote small displacements from these positions via $\delta\mathbf{R}_{\alpha p} = \mathbf{R}_{\alpha p}-\mathbf{R}_{\alpha p}^0$. 
Canonically conjugate momenta, $\mathbf{P}_{\alpha p}$, can then be defined by the commutation relation $[\mathbf{R}_{\alpha p},\mathbf{P}_{\alpha p}] = i\hbar\delta_{\alpha ,\alpha '}\delta_{p,p'}$.

Using the textbook definition of the interatomic force constant matrix and it's Fourier transform, the dynamical matrix (see \ref{section: SI Normal Mode Description}), we define the complex normal coordinates and momentum for a given phonon quasi-momentum $\mathbf{q}$ and branch $\nu$ as the following linear transformation :
\begin{equation}\label{eq:complex normal coordinates}
    \begin{split}
        z_{\mathbf{q}\nu} &= N_p^{-1/2}\sum_{\alpha p}e^{-i\mathbf{q}\cdot\mathbf{R}_p}(M_{\alpha}/M_0)^{1/2}\mathbf{e}^*_{\alpha\nu}(\mathbf{q})\cdot\delta\mathbf{R}_{\alpha p}\\
        P_{\mathbf{q}\nu} &= N_p^{-1/2}\sum_{\alpha p}e^{-i\mathbf{q}\cdot\mathbf{R}_p}(M_0/M_{\alpha})^{1/2}\mathbf{e}^*_{\alpha\nu}(\mathbf{q})\cdot\mathbf{P}_{\alpha p}.
    \end{split}
\end{equation}
Here, $\mathbf{e}_{\alpha\nu}(\mathbf{q})\in\mathbb{C}^d$ is the normal mode of vibration describing the displacement of atom $\alpha$ in the primitive cell with mass $M_{\alpha}$. $M_0$ is a reference mass, taken to be the mass of the proton. The inverse of this transformation reads as
\begin{equation}\label{eq:complex normal coordinate inverse}
    \begin{split}
        \delta\mathbf{R}_{\alpha p} &= N_p^{-1/2}\sum_{\mathbf{q}\nu}e^{i\mathbf{q}\cdot\mathbf{R}_p}(M_0/M_{\alpha})^{1/2}\mathbf{e}_{\alpha\nu}(q)z_{\mathbf{q}\nu}\\
        \mathbf{P}_{\alpha p} &= N_p^{-1/2}\sum_{\mathbf{q}\nu}e^{i\mathbf{q}\cdot\mathbf{R}_p}(M_{\alpha}/M_0)^{1/2}\mathbf{e}_{\alpha\nu}(q)P_{\mathbf{q}\nu}.
    \end{split}
\end{equation}
It is easily shown that the complex normal position and momenta obey the normal commutation relations for phonons \cite{Bruesch1982,Coleman2015,Giustino2017,Caruso_Zacharias2022}:
\begin{equation}
    [z_{\mathbf{q}\nu},P_{-\mathbf{q}'\nu'}] = i\hbar\delta_{\mathbf{q}\mathbf{q}'}\delta_{\nu\nu'}.
\end{equation}

The redundancy in the complex normal coordinates, seen by $z_{-\mathbf{q}\nu} = z_{\mathbf{q}\nu}^*$, can be removed by introducing the so called real normal coordinates \cite{Bruesch1982,Giustino2017}. To show how this is done, we start by partitioning the $\mathbf{q}$ grid in the first BZ into three sets, in the manner of Giustino and Br\"{u}esch. Call set $\mathcal{A}$ the set of vectors invariant under inversion modulo addition with a reciprocal lattice vector $\mathbf{G}$, i.e. $\mathbf{q} =-\mathbf{q} + \mathbf{G}$ . Set $\mathcal{B}$ and $\mathcal{C}$ are partitioned in such a way that all vectors $\mathbf{q}\in \mathcal{C}$ are obtained from $\mathbf{q}'\in \mathcal{B}$ via inversion, i.e. $\mathbf{q}'=-\mathbf{q} + \mathbf{G}$. This leads to the following definitions
\begin{equation}
\begin{split}
z_{\mathbf{q}\nu} = \begin{cases}
x_{\mathbf{q}\nu} \null\quad &\text{for}\ \mathbf{q}\in\mathcal{A}\\
x_{\mathbf{q}\nu} +iy_{\mathbf{q}\nu}\null\quad &\text{for}\ \mathbf{q}\in\mathcal{B}\\
\end{cases}
\end{split}
\end{equation}
\begin{equation}
\begin{split}
P_{\mathbf{q}\nu} = \begin{cases}
r_{\mathbf{q}\nu} \null\quad &\text{for}\ \mathbf{q}\in\mathcal{A}\\
r_{\mathbf{q}\nu} +is_{\mathbf{q}\nu}\null\quad &\text{for}\ \mathbf{q}\in\mathcal{B}\\
\end{cases}
\end{split}
\end{equation}whereby the following properties hold true
\begin{equation}\label{eq:normal coordinate inversion properties}
    \begin{split}
        x_{-\mathbf{q}\nu} &= x_{\mathbf{q}\nu},\null\quad r_{-\mathbf{q}\nu} = r_{\mathbf{q}\nu}\\
        y_{-\mathbf{q}\nu} &= -y_{\mathbf{q}\nu},\null\quad s_{-\mathbf{q}\nu} = -s_{\mathbf{q}\nu}\\
    \end{split}
\end{equation}
Now we can rewrite (\ref{eq:complex normal coordinate inverse}) as
\begin{equation}\label{eq:real space normal coordinates}
    \begin{split}
        \delta\mathbf{R}_{\alpha p} &= N_p^{-1/2}(M_0/M_{\alpha})^{1/2}\left( \sum_{\mathbf{q}\in \mathcal{A},\nu} \mathbf{e}_{\alpha\nu}(\mathbf{q})x_{\mathbf{q}\nu}\cos{\left(\mathbf{q}\cdot\mathbf{R}_p\right)}\right. \\
        &+ \left.2\text{Re} \left[\sum_{\mathbf{q}\in \mathcal{B},\nu} e^{i\mathbf{q}\cdot\mathbf{R}_p}\mathbf{e}_{\alpha\nu}(\mathbf{q})(x_{\mathbf{q}\nu} + iy_{\mathbf{q}\nu})\right]\vphantom{\int_1^2}\right)\\
         \mathbf{P}_{\alpha p} &= N_p^{-1/2}(M_{\alpha}/M_0)^{1/2}\left( \sum_{\mathbf{q}\in \mathcal{A},\nu} \mathbf{e}_{\alpha\nu}(\mathbf{q})r_{\mathbf{q}\nu}\cos{\left(\mathbf{q}\cdot\mathbf{R}_p\right)}\right. \\
         &+ \left.2\text{Re} \left[\sum_{\mathbf{q}\in \mathcal{B},\nu} e^{i\mathbf{q}\cdot\mathbf{R}_p}\mathbf{e}_{\alpha\nu}(\mathbf{q})(r_{\mathbf{q}\nu} + is_{\mathbf{q}\nu})\right]\vphantom{\int_1^2}\right),\\
    \end{split}
\end{equation}
where the sum over $\mathcal{C}$ has been included by taking twice the real part of the sum over $\mathcal{B}$. With this removal of the redundancy, one can see that this linear transformation contains $dN_cN_p$ independent variables corresponding to $x_{\mathbf{q}\nu}$ for $\mathbf{q}\in \mathcal{A}$ and $x_{\mathbf{q}\nu},y_{\mathbf{q}\nu}\in B$, with canonical conjugates $r_{\mathbf{q}\nu}$ for $\mathbf{q}\in \mathcal{A}$ and $r_{\mathbf{q}\nu},s_{\mathbf{q}\nu} $ for $ \mathbf{q}\in \mathcal{B}$, defining canonical commutation relations
\begin{equation}\label{eq:real normal coordinate commutation relations}
    \begin{split}
        \left[ x_{\mathbf{q}\nu},r_{\mathbf{q}'\nu'}\right] &= \left[ y_{\mathbf{q}\nu},s_{\mathbf{q}'\nu'}\right]= i\hbar\delta_{\mathbf{q}\mathbf{q}'}\delta_{\nu\nu'}\\
        \left[ x_{\mathbf{q}\nu},y_{\mathbf{q}'\nu'}\right] &= \left[ r_{\mathbf{q}\nu},s_{\mathbf{q}'\nu'}\right] = 0\\
        \left[x_{\mathbf{q}\nu},s_{\mathbf{q}'\nu'}\right] &= \left[y_{\mathbf{q}\nu},r_{\mathbf{q}'\nu'}\right] = 0.
    \end{split}
\end{equation}
Following Giustino we define the characteristic length scale of the phonon frequencies as
\begin{equation}\label{eq:phonon zero point length}
l_{\mathbf{q}\nu} = \begin{cases}
\left(\frac{\hbar}{2M_0\omega_{\mathbf{q}\nu}}\right)^{1/2}\null\quad\text{for}\null\quad \mathbf{q}\in \mathcal{B},\mathcal{C}\\
2\left(\frac{\hbar}{2M_0\omega_{\mathbf{q}\nu}}\right)^{1/2}\null\quad\text{for}\null\quad \mathbf{q}\in \mathcal{A}
\end{cases}
\end{equation}
and rescale the coordinates as $\tilde{z}_{\mathbf{q}\nu} = z_{\mathbf{q}\nu}/l_{\mathbf{q}\nu}$ and $\tilde{P}_{\mathbf{q}\nu} = P_{\mathbf{q}\nu}l_{\mathbf{q}\nu}/\hbar$. With this linear transformation we use the properties of the normal modes subject to the BvK boundary conditions to rewrite the real space nuclear Hamiltonian in the harmonic limit as:
\begin{equation}\label{eq:phonon hamiltonian real normal coordinates}
\begin{split}
    \hat{H}_{\text{ph}} &= \frac{1}{2}\sum_{\mathbf{q}\in \mathcal{A},\nu}\omega_{\mathbf{q}\nu}\left(\tilde{r}_{\mathbf{q}\nu}^2 + \tilde{x}_{\mathbf{q}\nu}^2\right) \\
    &+ \frac{1}{2} \sum_{\mathbf{q}\in \mathcal{B},\nu}\omega_{\mathbf{q}\nu}\left(\tilde{r}_{\mathbf{q}\nu}^2 + \tilde{x}_{\mathbf{q}\nu}^2 + \tilde{s}_{\mathbf{q}\nu}^2 + \tilde{y}_{\mathbf{q}\nu}^2\right).
\end{split}
\end{equation}

\subsubsection{Phonon Wigner function}
The Wigner transform of the density matrix of a set of uncoupled phonons in the canonical ensemble can be written in terms of the reduced coordinates as \cite{Hillery1984}: 
\begin{equation}\label{eq:phonon distribution}
\begin{split}
\rho_{\text{ph}, W}&= \\
\prod_{\nu,\mathbf{q}\in\mathcal{A,B}}&\frac{\text{tanh}(\beta\omega_{\mathbf{q}\nu}/2)}{\pi}
\exp\left[-\text{tanh}(\beta\omega_{\mathbf{q}\nu}/2)\left(\tilde{r}_{\mathbf{q}\nu}^2 + \tilde{x}_{\mathbf{q}\nu}^2\right)\right]\\
\times\prod_{\nu,\mathbf{q}\in\mathcal{B}}&\frac{\text{tanh}(\beta\omega_{\mathbf{q}\nu}/2)}{\pi}\exp\left[-\text{tanh}(\beta\omega_{\mathbf{q}\nu}/2)\left(\tilde{s}_{\mathbf{q}\nu}^2 + \tilde{y}_{\mathbf{q}\nu}^2\right)\right],
\end{split}
\end{equation}
for $\beta=1/k_bT$, where the reduced coordinates are now treated as continuous degrees of freedom. We use this distribution to sample the phonon modes when performing real space supercell calculations; the nuclear configuration associated with a particular phonon coordinate configuration is obtained by simply using Eq. (\ref{eq:real space normal coordinates}) after sampling the reduced coordinates from Eq. (\ref{eq:phonon distribution}). The required inputs are $\mathbf{e}_{\alpha\nu}(\mathbf{q})$ and $\omega_{\mathbf{q}\nu}$, which are easily obtained via Density Functional Perturbation Theory (DFPT) calculations such as the implementation in Quantum Espresso \cite{Giannozzi2009,Giannozzi2017}.

We note here that while sampling the equilibrium phonon distribution in Eq. (\ref{eq:phonon distribution}) exactly captures the quantum statistics of the phonon system only in the harmonic limit, the time evolution is not constrained to this limit as the nuclei are subject to the (Ehrenfest) forces coming from the driven electronic system. In this sense, the phonon coordinate picture can be viewed as a convenient basis in which the nuclear system can be initialized, rather than a limitation of MTEF to the harmonic approximation. 

\subsection{Connection to Static Displacement Methods}
In the special displacement method (SDM) introduced by Zacharias and Giustino \cite{Zacharias2016,Zacharias2020}, static thermodynamic properties of periodic systems (within the Born-Oppenheimer approximation) are calculated by taking specific phonon coordinate configurations from the Williams-Lax nuclear coordinate distribution, written in reciprocal space \cite{Zacharias2020}. For an arbitrary operator $\hat{O}$ the expectation value at a given temperature is given as
\begin{equation}\label{eq:ZG average}
\begin{split}
    O(T) = &\prod_{\nu,\mathbf{q}\in\mathcal{A}}\int \frac{dx_{\mathbf{q}\nu}}{\sqrt{\pi}\sigma_{\mathbf{q}\nu}}e^{-\frac{x_{\mathbf{q}\nu}^2}{\sigma_{\mathbf{q}\nu}^2}}\\
    \times&\prod_{\nu,\mathbf{q}\in\mathcal{B}}
    \int \frac{dx_{\mathbf{q}\nu}dy_{\mathbf{q}\nu}}{\pi\sigma^2_{\mathbf{q}\nu}}e^{-\frac{x_{\mathbf{q}\nu}^2+y_{\mathbf{q}\nu}^2}{\sigma_{\mathbf{q}\nu}^2}}O^{\{x_{\mathbf{q}\nu},y_{\mathbf{q}\nu}\}}(T),
\end{split}
\end{equation}
with $\sigma_{\mathbf{q}\nu}^2 = l^2_{\mathbf{q}\nu}\left(2n_{\mathbf{q}\nu,T} + 1\right)$, for the Bose-Einstein occupation of the mode, $n_{\mathbf{q}\nu,T} = \left[\exp\left(\beta\hbar\omega_{\mathbf{q}\nu}\right)-1\right]^{-1}$, and $O^{\{x_{\mathbf{q}\nu},y_{\mathbf{q}\nu}\}}(T)$ referring to the expectation value of $\hat{O}$ at temperature $T$ with respect to the electronic system, evaluated at phonon configuration $\{x_{\mathbf{q}\nu},y_{\mathbf{q}\nu}\}$.

The connection to the Wigner transformation of the phonon coordinates can been seen by inserting $\sigma_{\mathbf{q}\nu}$ and $n_{\mathbf{q}\nu,T}$ for the distribution functions of Eq. (\ref{eq:ZG average}). Keeping in mind our definition of $l_{\mathbf{q}\nu}$ in Eq. (\ref{eq:phonon zero point length}), one immediately arrives at Eq. (\ref{eq:phonon distribution}) for the position coordinates. Therefore, including the phonon momenta is in some sense an extension of this method to dynamic ions. Thus, given the success of the SDM in capturing phonon renormalization using small numbers of displacement samples, one can consider sampling approaches inspired by this method which incorporate momenta. We explore this possibility in \ref{section: SI Alternative Sampling Approaches}, but find no significant advantage for the observables studied here.

In what follows, we refer to a simulation protocol as `static' when we sample exclusively positions from Eq. (\ref{eq:phonon distribution}) (i.e. from the distribution function in Eq. (\ref{eq:ZG average})), constrain the phonon coordinates at this initial configuration and evolve only the electronic system in time. This approximation is sometimes referred to as the `clamped ion', `frozen phonon', or `static disorder' approximation. In contrast we refer to a protocol as `dynamic' when both phonon coordinate and momenta initial conditions are sampled from Eq. (\ref{eq:phonon distribution}) and the phonon coordinates evolve along with the electronic system according to Eq. (\ref{eq:MTEF eom}). 

\subsection{Time-dependent Boltzmann equation}
In order to make connections with statistical mechanics approaches based on perturbation theory, we also make comparisons with a time-dependent Boltzmann equation treatment of the problem. Starting from Fermi's golden rule for the first-order rate equation for electron occupation in band $n$ at crystal momentum $\mathbf{k}$, $f_{n\mathbf{k}}$ and phonon occupation $n_{\mathbf{q}\nu}$ due to electron-phonon scattering processes:
\cite{Caruso2022,Bernardi2016}
\begin{equation}
\begin{split}
    \partial_t f_{n\mathbf{k}} &= 2\pi \sum_{m\nu\mathbf{q}}|g_{mn}^{\nu}(\mathbf{k},\mathbf{q})|^2\\
    \times \{&(1-f_{n\mathbf{k}})f_{m\mathbf{k}+\mathbf{q}}\delta(\epsilon_{n\mathbf{k}} - \epsilon_{m\mathbf{k}+\mathbf{q}} + \omega_{\mathbf{q}\nu})(n_{\mathbf{q}\nu}+1)\\ 
    +&(1-f_{n\mathbf{k}})f_{m\mathbf{k}+\mathbf{q}}\delta(\epsilon_{n\mathbf{k}} - \epsilon_{m\mathbf{k}-\mathbf{q}} - \omega_{\mathbf{q}\nu})n_{\mathbf{q}\nu}\\
    +&f_{n\mathbf{k}}(1-f_{m\mathbf{k}+\mathbf{q}})\delta(\epsilon_{n\mathbf{k}} - \epsilon_{m\mathbf{k}-\mathbf{q}} - \omega_{\mathbf{q}\nu})(n_{\mathbf{q}\nu}+1)\\
    +&f_{n\mathbf{k}}(1-f_{m\mathbf{k}+\mathbf{q}})\delta(\epsilon_{n\mathbf{k}} - \epsilon_{m\mathbf{k}-\mathbf{q}} + \omega_{\mathbf{q}\nu})n_{\mathbf{q}\nu}
    \},
\end{split}
\end{equation}
and
\begin{equation}
\begin{split}
    \partial_t n_{\mathbf{q}\nu}  = 4\pi \sum_{mn\mathbf{k}}&|g_{mn}^{\nu}(\mathbf{k},\mathbf{q})|^2f_{n\mathbf{k}}(1-f_{m\mathbf{k}+\mathbf{q}})\\
    \times \{&\delta(\epsilon_{n\mathbf{k}} - \epsilon_{m\mathbf{k}+\mathbf{q}} - \omega_{\mathbf{q}\nu})(n_{\mathbf{q}\nu}+1) \\
    - &\delta(\epsilon_{n\mathbf{k}} - \epsilon_{m\mathbf{k}+\mathbf{q}} + \omega_{\mathbf{q}\nu})n_{\mathbf{q}\nu}\}.
\end{split}
\end{equation}
Here $g_{mn}^{\nu}(\mathbf{k},\mathbf{q})$ are the electron-phonon matrix elements in the band basis. Since we focus mainly on the short time scale dynamics, we do not include phonon-phonon scattering in our TDBE analysis. 

\section{Tight Binding Model}
We treat the electronic structure of hBN in a widely used DFT-based tight binding model with nearest neighbor hopping between the two inequivalent sublattice sites. While, in principle, the electron-phonon coupling matrix elements can be extracted from existing DFPT packages, these quantities are almost always given as absolute values which can be used in the TDBE. As the time evolution of the electronic dynamics in MTEF also requires the complex phase of the coupling matrix elements, we derive a BZ extensive expression for the coupling. We treat the nuclear dependence of the electronic Hamiltonian by modeling the hopping term to be exponentially dependent on interatomic distance: \begin{equation}t(\mathbf{R}_i,\mathbf{R}_j)=t_0\exp\left(-b\left[\frac{|\mathbf{R}_i-\mathbf{R}_j|}{d_0} - 1\right]\right), \end{equation} where $t_0$ is the equilibrium hopping term, $b$ is the electron-phonon coupling factor, and $d_0$ is the equilibrium distance between sublattice sites $i$ and $j$. This fitting is common in literature for graphene tight binding models, and $b$ can be related to experimental observables \cite{Pereira2009,Droth2016,Mohanty2019}. 

For simplicity we restrict our study to two dimensions by only considering phonon branches with no out of plane component. We perform a second order expansion of ionic displacement from equilibrium while restricting the hopping term expansion to first order, ignoring the second order electron-phonon coupling coefficients, or Debye-Waller terms, which is quite often done in literature \cite{Giustino2017,Mohanty2019}.  By rewriting the electronic operators in terms of plane waves and the nuclear displacements in terms of phonon complex normal coordinates, we obtain the following reciprocal space Hamiltonian (see \ref{section: SI EPh-derivation} for details): 
\begin{equation}\label{eq:Hamiltonian reciprocal space}
\begin{split}
&\hat{H}_W(\mathbf{X}) = \frac{1}{2}\sum_{\mathbf{q}\nu} \omega_{\mathbf{q}\nu}\left(\tilde{P}_{\mathbf{q}\nu}^2 + \tilde{z}_{\mathbf{q}\nu}^2\right) + \sum_{\alpha\in\{a,b\}}\Delta_{\alpha}\hat{\alpha}_{\mathbf{k}}^{\dagger}\hat{\alpha}_{\mathbf{k}}\\
&-t_0\sum_{\mathbf{k}} \left(\hat{a}_{\mathbf{k}}^{\dag}\hat{b}_{\mathbf{k}}\sum_{\boldsymbol{\delta}}e^{i\mathbf{k}\cdot\boldsymbol{\delta}}+ c.c.\right) +\sum_{\mathbf{q}\nu} \tilde{z}_{\mathbf{q}\nu}l_{\mathbf{q}\nu}\hat{M}(\mathbf{q},\nu),\\
&\hat{M}(\mathbf{q},\nu) =  \sum_{\mathbf{k}\boldsymbol{\delta}}g_{\nu}^{\delta}(\mathbf{q})\left(\hat{a}_{\mathbf{k}+\mathbf{q}}^{\dag}\hat{b}_{\mathbf{k}}e^{i\mathbf{k}\cdot\boldsymbol{\delta}} + \hat{b}_{\mathbf{k}}^{\dag}\hat{a}_{\mathbf{k}-\mathbf{q}}e^{-i\mathbf{k}\cdot\boldsymbol{\delta}}\right).\\
\end{split}
\end{equation}
Here $\mathbf{X} = (\mathbf{z},\mathbf{P})$ is the collection of phonon coordinates, $\hat{\alpha}_{\mathbf{k}}$ are the electronic site operators for sites $\alpha=\{a,b\}$ with onsite energies $\Delta_{\alpha}=\pm \Delta$ responsible for opening the gap $E_g=2|\Delta|$. Nearest neighbor sites are connected by vectors $\boldsymbol{\delta}$. The matrices $\hat{M}(\mathbf{q},\nu)$ describe the coupling of the phonon normal coordinates to the electronic system, resulting in the scattering of electronic planewaves from $\mathbf{k}\to\mathbf{k}\pm\mathbf{q}$, and depend on coupling terms $g_{\nu}^{\delta}(\mathbf{q})$, which themselves depend on $\mathbf{e}_{\alpha\nu}(\mathbf{q})$.

The required inputs for this model are the lattice constant $a_0$, onsite energies $\Delta_{\alpha}$, hopping term $t_0$, phonon dispersion $\omega_{\mathbf{q}\nu}$,  polarization $\mathbf{e}_{\nu}(\mathbf{q})$ and electron-phonon coupling factor $b$. In this work, we obtained the lattice constant and phonon information from Quantum Espresso cell relaxation and phonon dispersion calculations, and fit the hopping term, onsite energies and coupling factor from a symmetric two band approximation to the conduction band calculated using an uncorrected LDA xc functional. See the Computational Methods section for further details. 

The present tight-binding model clearly involves a major simplification, as it treats the electronic system at the independent particle level. However, this simplified treatment is not entirely unreasonable as one can eliminate excitonic effects in experimental setups by placing the monolayer onto a conductive substrate, which screens the electric field and allows the study of free carriers \cite{Cabo2015,Ulstrup2017}. Furthermore, the purpose of the TB model is mainly in developing comparisons with \textit{ab initio} simulations, which go beyond this limitation. Nevertheless, it would be interesting to go beyond this limit in future applications.

\subsubsection{Implementing the MTEF Method with the TB Model} 
For each initial condition $\mathbf{X}^0$ that is sampled from $\rho_{\text{ph},W}(\mathbf{X})$, one can find the single particle orbitals for $\hat{H}_W(\mathbf{X}^0)$, 
\begin{equation}\label{eq:electronic-orbs}    \hat{H}_W(\mathbf{X}^0)\ket{\psi_l(\mathbf{X}^0)} = \epsilon_{l}(\mathbf{X}^0)\ket{\psi_l(\mathbf{X}^0)}\\
\end{equation}and construct the density operator,
\begin{equation}\label{eq:electronic-init}
    \hat{\rho}_{\text{e}} = \sum_{l}f(\epsilon_l^{0},T)\ket{\psi_l}\bra{\psi_l},
\end{equation}
where $f(\epsilon_l^{0},T)$ is the Fermi occupation at temperature $T$ evaluated for the orbital energy $\epsilon_l(\mathbf{X}^0) = \epsilon_l^{0}$ at this initial configuration. In the above expression for the electronic density we have suppressed the dependence of the orbital on the phonon coordinates. One may then propagate a set of trajectories, associated with the set of initial conditions, according to the MTEF equations of motion (Eq.(2)), which in the case of the present TB model are more conveniently expressed as follows, 
\begin{equation}
\begin{split}
\dot{\tilde{P}}_{\mathbf{q}\nu}^i &= -\left(\omega_{\mathbf{q}\nu}\tilde{z}_{\mathbf{q}\nu}^i + l_{\mathbf{q}\nu}\text{Tr}\left[M_{\mathbf{q}\nu}\rho^i_e(t)\right]\right)\\
\dot{\tilde{z}}_{\mathbf{q}\nu}^i &= \omega_{\mathbf{q}\nu}\tilde{P}^i_{\mathbf{q}\nu}\\
i\partial_t\ket{\psi_l^i} &= \hat{H}_W(\mathbf{X}^i(t))\ket{\psi_l^i}.\\
\end{split}
\end{equation} 

We have validated the accuracy of the MTEF treatment of this form of tight binding model by comparing with the nonequilibrium Green's function (NEGF) results of Nery and Mauri \cite{Nery2022} for the phonon renormalization of the electronic spectral function in graphene, and we find that the spectral function generated from MTEF simulations is in good agreement with the NEGF results. See the \ref{section: SI graphene} for details. 

\begin{figure*}
    \includegraphics[width=\textwidth]{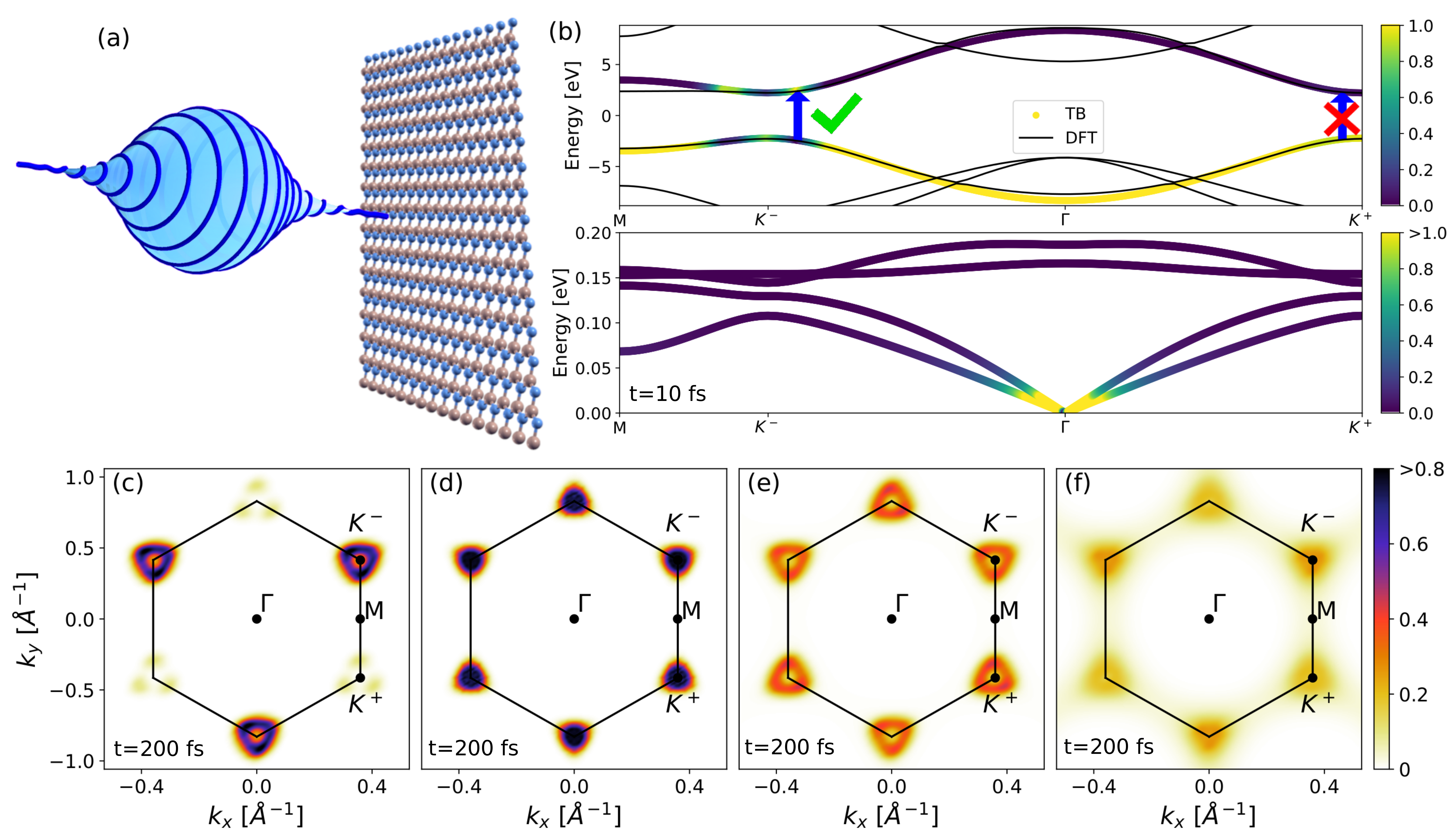}
    \caption{Panel (a) shows a schematic of circularly polarized light exciting a sheet of hBN. The upper half of panel (b) shows the electron bands calculated using DFT with an LDA xc functional, alongside the tight binding (TB) bands fit to match the band onset. The Fermi level has been centered between the conduction and valence bands. The arrow indicates the approximate energy of excitation around a specific $K$ valley, and the shading indicates the occupation of states in the TB model just after the laser pulse, interpolated from the MTEF results. The lower half of panel (b) shows the phonon dispersion calculated using DFPT, with shading indicating the initial phonon occupation number at 300 K, with the translational modes at $\Gamma$ set to zero. The bottom four panels show the occupation of the conduction band 200 fs after the laser pulse under different approximations to the electron-phonon dynamics: (c) initially equilibrium geometry with zero velocity, (d) TDBE, (e) static displacement, and (f) MTEF dynamics. The phonon system in panels (d-f) is sampled at/set to 300 K. The labeling of the otherwise degenerate $K^+/K^-$ points is based off the polarization of the pump and the scale is capped at 0.8, to emphasize small differences.}
    \label{fig:occupation-4panel}
\end{figure*}

\section{Results}
\subsection{Excited Carrier Relaxation in hBN}\label{sec: excited carrier relaxation}
Due to the lack of inversion symmetry, hBN is a strong insulator with a gap around the $K/K'$ points that is depicted in panel (b) of Fig. \ref{fig:occupation-4panel}. 

This also leads to the potential for selective excitation in the BZ around the $K/K'$ points upon irradiation with circularly polarized light. 

For simplicity, we did not include direct coupling between the laser field and the phonon modes, and as such the laser field is only indirectly coupled to the phonons via the modification of the electronic occupations. To couple the electronic system to the laser field in the long-wave approximation we use a Peierls substitution to modify the electronic wavevector $\mathbf{k}$:
\begin{equation}\label{eq:k(t)}
\mathbf{k}(t) = \mathbf{k} - \int^t \mathbf{E}(t')dt' = \mathbf{k} + \frac{1}{c}\mathbf{A}(t).
\end{equation}

We use a circularly polarized laser pulse $\mathbf{A}(t)$ using a $\cos^2$ envelope with polarization defined by:
\begin{equation}\label{eq:laser}
\begin{split}   
    \mathbf{A}(t) &= \begin{cases}
        \mathbf{A}_0(t)\cos^2\left(\frac{\pi}{T_{\text{pump}}}(t-T_{\text{pump}}/2)\right)&,\ t<T_{\text{pump}}\\
        \null\qquad0 &\text{, else}
    \end{cases}\\
    \mathbf{A}_0(t) &= \frac{A_0}{\sqrt{2}}\left(\text{Re}\left[e^{is\omega t}\right]\hat{\mathbf{x}} + \text{Im}\left[e^{is\omega t}\right]\hat{\mathbf{y}}\right).
\end{split}
\end{equation}

We utilize a pump pulse duration of ten optical cycles at carrier frequency $\omega=5$ eV, such that $T_{\text{pump}}\approx 8.3$ fs (FWHM $\approx 4.1$ fs), with amplitude $A_0=5$ a.u. corresponding to a peak intensity of $7.88\times 10^{11}$ W/cm$^2$. The handedness of the laser is controlled by $s=\{1,-1\}$, creating left and right circularly polarized light, and we distinguish the $K$ points by which sign of $s$ excites them as $K^+/K^-$.

After exposure to the pump pulse we track the occupation of the conduction band states, $f_{c\mathbf{k}} = |c_{\mathbf{k}}|^2$, by taking the expectation value of the projector:
\begin{equation}
    c_{\mathbf{k}}(t) = \frac{1}{N_t}\sum_{i}^{N_t}\text{Tr}\left[\hat{\rho}_i(t)\ket{c\mathbf{k}}\bra{c\mathbf{k}}\right].
\end{equation} Strictly speaking this projector is only valid when the laser field is turned off, otherwise one must project onto the Houston states $\ket{c\mathbf{k}(t)}$ \cite{Houston1940, Krieger1986}. For simplicity we project onto the equilibrium band states, and indicate when the laser field is on (and therefore where this measure is only approximate) using grey shading in the background. 
In panels (c-f) of Fig. \ref{fig:occupation-4panel} we show a snapshot of the conduction band occupations taken 200 fs after the circularly polarized pump pulse, using the different approaches mentioned above. In all cases the system is initialized at 300 K. In Fig. \ref{fig:occupation-4panel} (c) we initialize the ionic system at the equilibrium geometry with zero velocity. The excited population established after the pulse has a large imbalance between the $K^-$ and $K^+$ valley carrier distributions and remains completely static throughout the propagation in this case due to a lack of decay channels. The marginal population around $K^+$, mostly around the $K^{\pm}M$ lines, and the dip at $K^-$ is due to pumping slightly above the gap, as seen in Fig. \ref{fig:occupation-4panel} (b), though there is no population at the symmetry forbidden $K^+$ point itself.  

Often in literature, a dynamics simulation will be referred to as Ehrenfest if the ions are allowed to move according to mean field forces, without a unique specification of their initial conditions. We show that performing the simulation with fixed ions or with dynamic ions starting with zero initial velocity, but in either case starting from the equilibrium geometry, results in no qualitative change in the excited electron occupation. By breaking the symmetries of the equilibrium geometry and including static disorder in the phonon system, there is a fundamental difference in the evolution of the excited electronic system.

The results of TDBE dynamics are shown in panel (d) of Fig. \ref{fig:occupation-4panel}. The initial carrier distribution rapidly equalizes between the $K^-/K^+$ valleys, while simultaneously contracting towards the lowest energy states available in the valley bottoms. These dynamics are restricted exclusively to relaxation of the electronic occupations because, as seen in panel (b), the optical phonons initially have no energy available to contribute to scattering the electronic system uphill in energy.

For the case of static disorder, shown in Fig. \ref{fig:occupation-4panel}(e), the valley occupations are effectively equalized by 200 fs, however there is no change in occupied energy levels, as seen by the ring around $K^{\pm}$ which is present in the occupation immediately after excitation as in panel (c). This can be explained in the framework of band folding: displacing the ionic positions in a supercell is equivalent to folding over the primitive cell BZ bands. This allows the excited charge carriers at $K^-$ to have a decay channel into the energetically degenerate folded $\mathbf{k}$ points around $K^+$. However, this is a strictly elastic scattering process and therefore is effectively restricted to states inside the initially excited energy window.

The results of allowing for dynamic ion motion at the MTEF level are shown in Fig. \ref{fig:occupation-4panel} (f). Here we see that in addition to homogenization of valley population, there is also some scattering of carriers out of the energetic window in which they were initially excited. One can see that in addition to scattering downwards in energy towards the $K^{\pm}$ points, there is also excitation of the electrons upward in the valleys along the $K^{\pm}M$ lines. Without exact numerical results it is difficult to address the accuracy of this phenomena, however it could be related to the ZPE leakage of the mean-field dynamics: although the optical phonon occupation is approximately zero initially, over time the ZPE of phonons near the gamma point is drained into the electronic system, incrementally exciting the charge carriers beyond what one sees in the TDBE results at this temperature. However, note that when propagating without pumping the electronic system, there is no loss of phonon ZPE and no heating of the electronic system as the band gap forbids the promotion of electrons due to the order of magnitude smaller phonon energies. See \ref{section: SI ZPE Leakage} for further details.

\begin{figure}
    \centering
    \includegraphics[width=\linewidth]{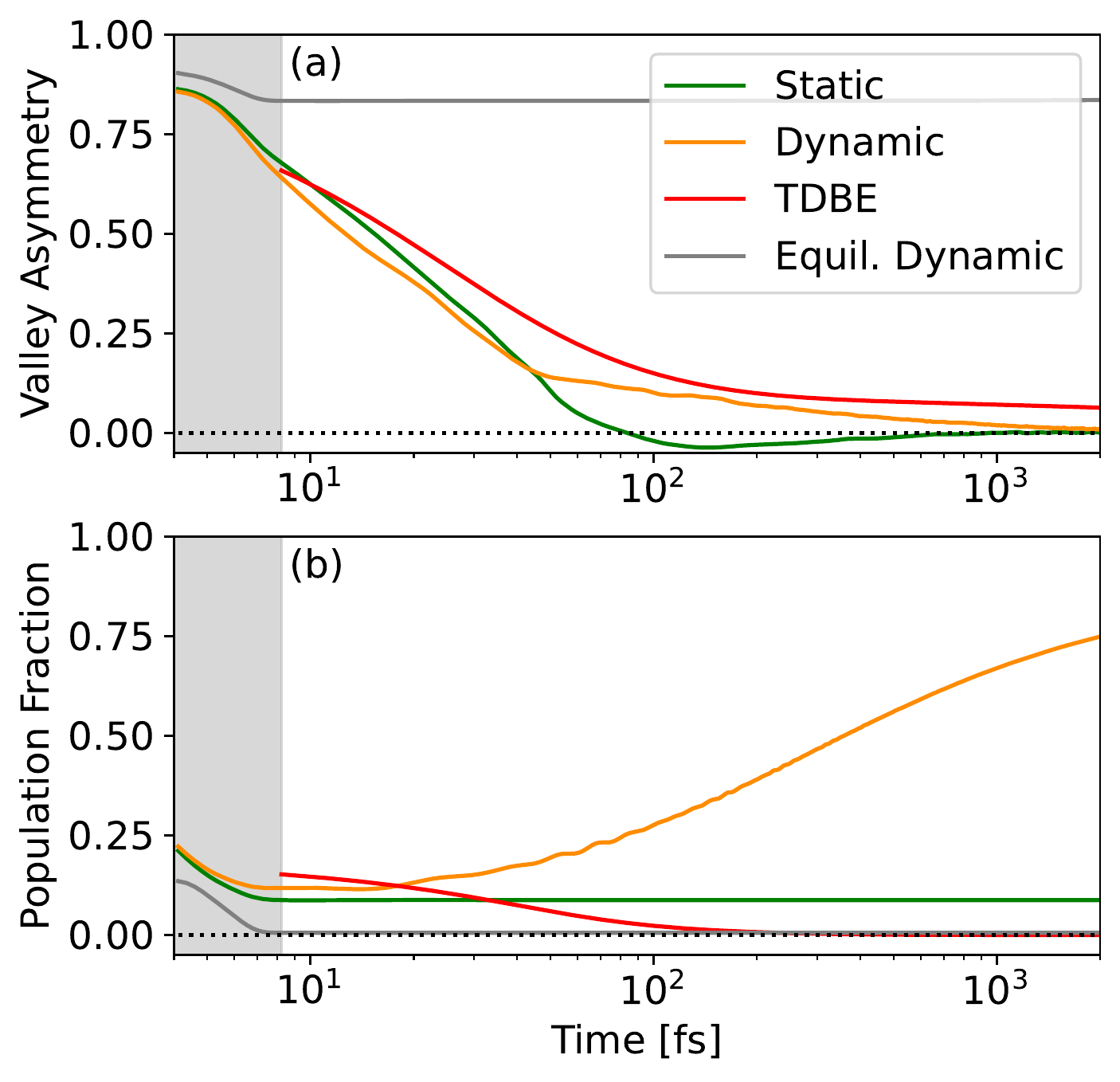}
    \caption{(a): Valley population asymmetry at $T=300$ K defined by Equation (\ref{eq: VP from occupation}). Panel (b) shows the fraction of conduction band population outside the $K^{\pm}$ regions. The grey region corresponds to when the pump is turned on and the dotted line at zero is a guide to the eye. 
    }
    \label{fig:valley-polarization-from-occupation}
\end{figure}
To assess the relaxation timescales in more detail, we can track the flow of excited charge carriers throughout the BZ by integration of the conduction band occupation within the populated region $A_{K^{\pm}}$ around the $K^{\pm}$ points:
\begin{equation}\label{eq: VP from occupation}
    \text{VA}_{\text{occ}}(t) = \frac{\int_{A_{K^-}} d^2\mathbf{k}\ f_{c\mathbf{k}}(t) - \int_{A_{K^+}} d^2\mathbf{k}\ f_{c\mathbf{k}}(t)}{\int_{A_{K^-}} d^2\mathbf{k}\ f_{c\mathbf{k}}(t) + \int_{A_{K^+}} d^2\mathbf{k}\ f_{c\mathbf{k}}(t)}.
\end{equation}
In the valleytronics literature this measure has been used to define the `valley polarization' or `valley asymmetry' \cite{Liu2019,Xu2021,Jimenez-Galan20} and (for TMDs) has been shown by perturbation theory to be strongly influenced by electron-phonon coupling \cite{Lin2022}. For our purposes we use this term to refer strictly to the population imbalance between the $K^{\pm}$ valleys, without reference to the spin resolved bands typically associated. We choose the integration regions to be $|\mathbf{k}-K^{\pm}|<0.36$\AA$^{-1}$, roughly corresponding to the populated areas Fig. \ref{fig:occupation-4panel}(c). In Fig. \ref{fig:valley-polarization-from-occupation}(a), we see that static displacement, MTEF dynamics and TDBE all capture an extremely rapid population rearrangement within the first $50$ fs. In the static displacement method there is a slight inversion of polarization which slowly decays over a $50$ fs-$2$ ps timescale, and the total population outside the valley regions in Fig. \ref{fig:valley-polarization-from-occupation}(b) remains fixed due to carrier energies being confined to their initial excitation energy window. Small changes to the radius of integration do not change the characteristics of these plots due to the normalization with respect to region population in Eq. (\ref{eq: VP from occupation}), and simply changes the quantitative values in Fig. \ref{fig:valley-polarization-from-occupation}(b).

The TDBE and MTEF dynamics clearly display a second time scale, starting from about $50$ fs in the case of MTEF and about $100$ fs for TDBE, where the valley equalization slows into an asymptotic-like behavior when the population imbalance dips below 0.15. Simultaneously, in the TDBE results there is a down-scattering of carriers which were initially outside of the valley regions as the excited electronic population emits energy into the phonon bath and assumes a more thermal distribution. In contrast, MTEF shows the presence of up-scattering processes as the ZPE from the phonon system is absorbed by the electronic system due to the classical nature of the approximation. See \ref{section: SI ZPE Leakage} for further details.

\begin{figure}
    \centering
    \includegraphics[width=\linewidth]{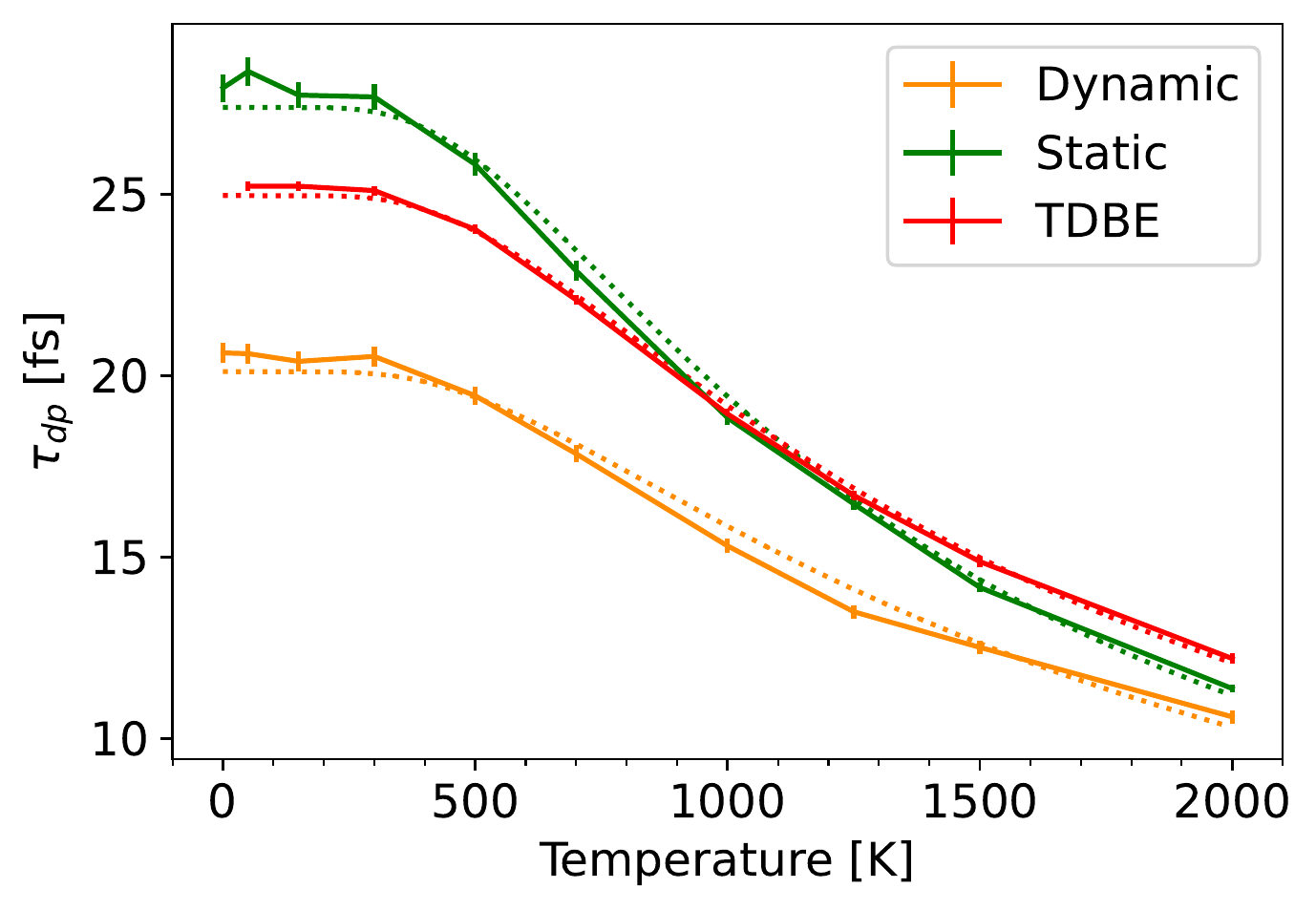}
    \caption{The timescale of an exponential fit of the valley depolarization rates calculated with Eq. (\ref{eq: VP from occupation}) over the first 50 fs, with an inverse scattering rate $\gamma$ fit to these data shown as the dotted line.
    }
    \label{fig:Depolarization-time-scale}
\end{figure}
We gain further insight to the effect of the phonon bath on the ultra-fast excited carrier relaxation or `valley depolarization' by varying the initial temperature of the system. In Fig. \ref{fig:Depolarization-time-scale} we show the characteristic time scales of depolarization obtained from fitting the valley asymmetry in the first 50 fs to an exponential decay function, $f(t)=a_0 + a_1\text{exp}\left(-t/\tau_{\text{dp}}\right)$, across multiple temperatures. The most apparent feature is an effective independence of the depolarization rate on temperature until about $300$K which can be easily attributed to the high phonon energies in hBN, meaning that the optical phonon branches only begin to have significant occupation at higher temperatures. This behavior has been experimentally observed in the steady-state photoluminescence polarization of MoS$_2$, which is directly proportional to the valley lifetime \cite{Zeng2012}, as well as in valley asymmetry decay times in WSe$_2$ measured by time-dependent Kerr rotation measurements \cite{Molina-Sanchez2017}. 

The depolarization timescales $\tau_{\text{dp}}$ can be related to the average phonon occupation via a linear function of the scattering rate: $\gamma = 1/\tau_{\text{dp}}$ where $\gamma = \gamma_0 + \alpha\braket{n_{\text{ph}}}$. Using the fit $\tau_{\text{dp}}$ values, we further fit the scattering rates by taking at each temperature the expected occupation value of the Bose-Einstein distribution at the average optical phonon energy, $0.16$ eV, showing very good agreement with the data. The fit scattering parameters are $\alpha_{\text{Dynamic}} = 0.072\ \text{fs}^{-1}$, $\alpha_{\text{Static}} = 0.081\ \text{fs}^{-1}$, and $\alpha_{\text{TDBE}} = 0.065\ \text{fs}^{-1}$. Although the ZPE leakage present in the MTEF dynamics can affect electronic populations at long time scales, these results show that for short time scales the MTEF decay rates broadly agree with the TDBE and static displacement approaches; all methods predict this timescale to depend on thermal activation of the phonon optical modes in a consistent manner.

In the context of the MTEF and static disorder simulations, it is worth pointing out that these results converge with an extremely small number of trajectories. For example, while the data shown in Fig. \ref{fig:valley-polarization-from-occupation} was obtained with $N_t=380$ trajectories, we can reproduce a graphically converged result with high probability through as little as two random samples. See \ref{section: SI Convergence} for further analysis. 

\subsection{Tracking Valley Asymmetry with Transient Circular Dichroism}\label{sec: transient circular dichroism}
\begin{figure*}
    \centering
    \includegraphics[width=\textwidth]{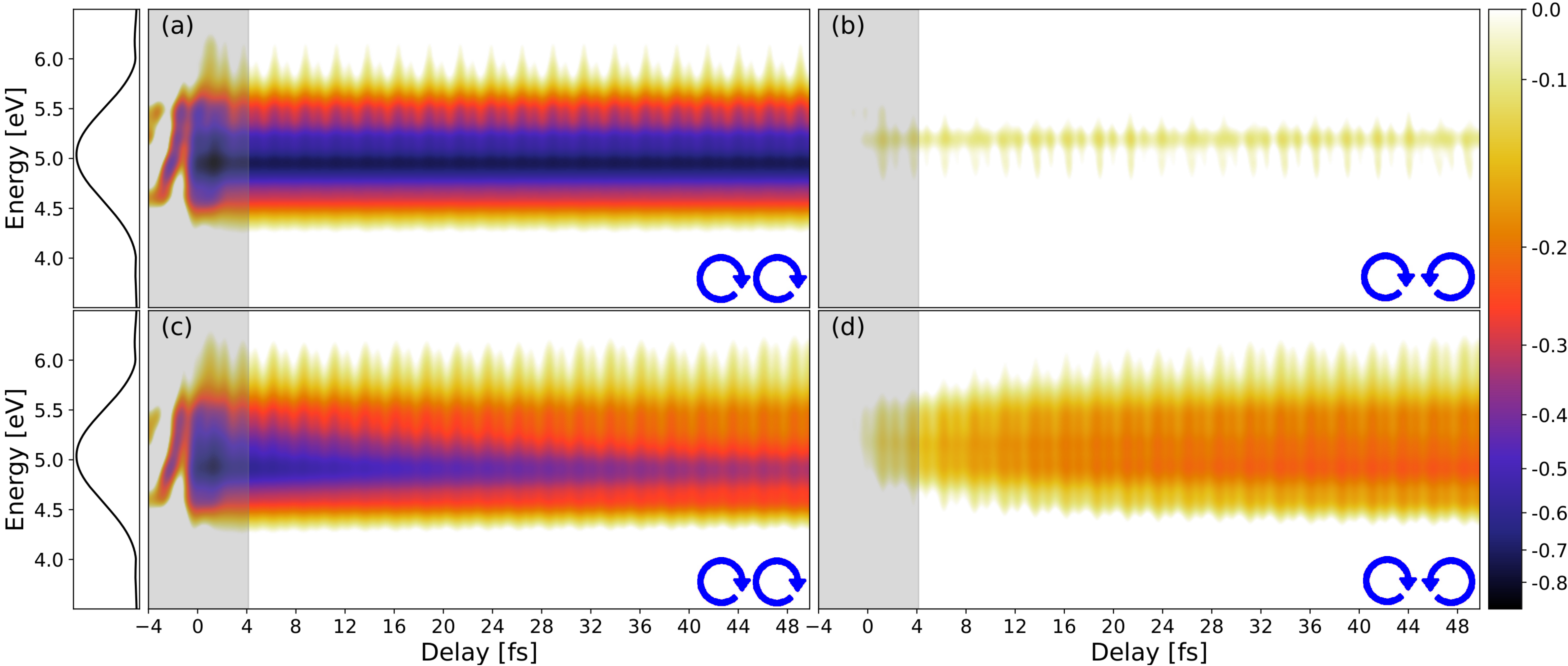}
    \caption{The real part of the TOC: $\text{Re}\left[\sigma(\omega,\tau)\right]$ calculated in the TB model. The arrows indicate the chirality of the pump and probe, either pumping and probing with the the same handedness or opposite handedness. Panels (a) and (b) are the equilibrium geometry results and the MTEF results at 300 K are shown in panels (c) and (d). The left most panels show the spectral weight of the pump, $|E(\omega)|$ which is active during the grey highlighted time span. The logarithmic scale has been set to emphasize the signal in the cross valley data, while minimizing the fluctuations arising from CEP locking.
    }
    \label{fig:tas-four}
\end{figure*}
Direct experimental observations of time dependent valley populations has been done in TMDs by performing time resolved measurements, such as time and angle resolved photoemission \cite{Ulstrup2017,Cabo2015}, time dependent Kerr rotation spectroscopy \cite{Molina-Sanchez2017}, helicity resolved two-dimensional electronic spectroscopy \cite{Lloyd2021}, and TAS \cite{Berghauser2018,Mai2014}. In this section we focus on TAS using circularly polarized light, Transient Circular Dichroism  (TCD).

Pumping the system with circularly polarized light produces an electronic current, $\mathbf{j}_{\text{pump}}(t)$, that depends on the polarization of the pump. Next, one probes the system with a much weaker circularly polarized probe pulse with a time delay between it's envelope center and the pump center of $\tau$ and a duration of $T_{\text{pr}}$ to generate the pump-probe current $\mathbf{j}_{\text{pump-probe}}(t,\tau)$, which encodes the excitation of the system induced by the pump at this delay, and depends on the polarization of the probe. Taking the difference between these currents \begin{equation}\label{eq: TAS current}
    \mathbf{j}_{\text{TAS}}(t,\tau) = \mathbf{j}_{\text{pump-probe}}(t,\tau) - \mathbf{j}_{\text{pump}}(t),
\end{equation}
allows one to calculate the intermediary transient optical conductivity (TOC) \cite{Sato2014}:
\begin{equation}\label{eq: intermediary TOC}
\begin{split}
    \tilde{\sigma}_{ij}(\omega,\tau) &= \frac{\int_{\tau-T_{\text{pr}}/2}^{T_f+\tau+T_{\text{pr}}/2} dt\ W(t/T_f)j^i_{\text{TAS}}(t,\tau)e^{i\omega t}}{\int_{\tau-T_{\text{pr}}/2}^{T_f+\tau+T_{\text{pr}}/2} dt\ W(t/T_f)E^j_{\text{probe}}(t)e^{i\omega t}}.
\end{split}
\end{equation} Here $i,j$ are the cartesian directions of the pump and probe, and we have inserted the mask function $W(x)=\exp\left(-\kappa x\right),\ \kappa = -\ln\left(10^{-3}/T_f\right)$ to damp the integrand of the Fourier transform, given the finite propagation time $T_f$. The trace of the optical conductivity is used throughout, $\sigma(\omega) = \text{Tr}\left[\sigma_{ij}(\omega)\right]$.
Finally by comparing the difference to the response of the system with no pump pulse but the same probe, $\sigma_{\text{no-pump}}(\omega)$ we obtain the true TOC:
\begin{equation}\label{eq: final TOC}
    \sigma(\omega,\tau) = \tilde{\sigma}(\omega,\tau) - \sigma_{\text{no-pump}}(\omega).
\end{equation}
As with the pump pulse, we also use a $\cos^2$ envelope for the probe, with a strength of $A_0=0.01$ a.u. at the same frequency as the pump for a single optical cycle, $T_{\text{pr}}\approx 0.8$ fs. The carrier envelope phase (CEP) between the pump and probe for each delay is fixed to zero. See \ref{section: SI Current operator} for the explicit formulas for the electronic current operator in the TB model and section \ref{section: MTEF-TAS} for the MTEF simulation protocol.

The results for the real component of the TOC calculated in the TB model are shown in Fig. \ref{fig:tas-four}, with a delay spacing of $\Delta \tau=0.5$fs and a propagation time of $T_f=30$fs. The pump chirality in all cases is $s_{\text{pump}}=-1$, the same used in the results of Fig. \ref{fig:occupation-4panel}, and the chirality of the probe is $s_{\text{probe}}=-1$ in the left column and $s_{\text{probe}}=+1$ in the right column. The periodic fluctuation of the signal is due to the CEP locking, and can in principle be removed if desired by averaging the results over several CEPs \cite{Walkenhorst2016,Bonafe2020}. Starting with the static equilibrium geometry results on the top row, in panel (a), we see that after the pump, at around $\tau=4$fs, the signal is saturated around the pumping frequency, whereas in panel (b) following the same pump, but with a probe pulse of the opposite chirality, there is a minimal response indicating a very small population slightly above $5$eV. By directly comparing to Fig. 
\ref{fig:occupation-4panel}(c) we see that the signal in Fig. \ref{fig:tas-four}(b) corresponds to the small population around (and energetically above) $K^+$ induced by the pump, while the saturated signal in Fig. \ref{fig:tas-four}(a) corresponds clearly to the large population in the $K^-$ valley. The TOC signal is of course unchanging after the pulse due to the lack of decay channels, again corresponding to the equilibrium results shown in Fig. \ref{fig:valley-polarization-from-occupation}. Turning to the MTEF results, we see a sharp attenuation of the signal in panel (c) following the pump, as well as a broadening of the range of the signal with respect to panel (a) due to scattering within and outside the $K^-$ valley. Concurrent with the attenuation of the $K^-$ valley signal around 5 eV there is a buildup of signal in panel (d) indicating a buildup of carrier density in the $K^+$ region. 

\begin{figure}
    \centering
    \includegraphics[width=\linewidth]{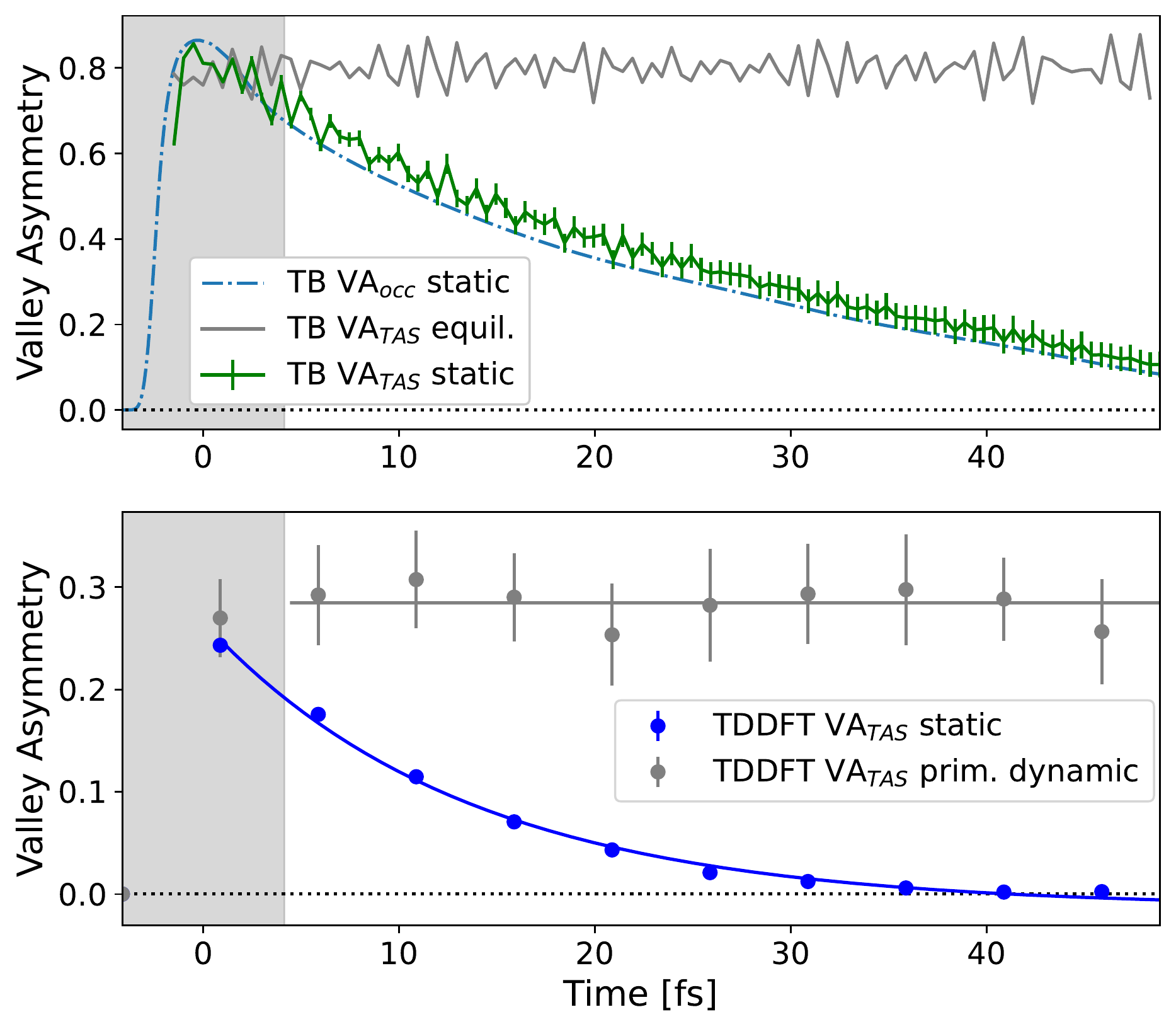}
    \caption{Extracting Valley Asymmetry from TAS calculations via Eq. (\ref{eq: VP from TAS}). The results of the same calculation done through TDDFT are also included. The grey shaded region is the time period in which the laser field is active. The characteristic decay times fit to the TAS data via an exponential starting from time $0$ are $\tau_{\text{dp}}=32\pm 2$fs for the tight binding calculation and $\tau_{\text{dp}}=13.5\pm 0.8$fs for the TDDFT calculation. The exponential fit to the TDDFT data is also plotted, with the grey line being a guide to the eye.}
    \label{fig:TAS-depol}
\end{figure}

We can generalize these predictions of the TOC to the calculated conduction band populations without reliance on defining valley integration areas by taking the difference of the columns shown in in Fig. \ref{fig:tas-four}(d) and (c), and integrating over the energy axis:
\begin{equation}\label{eq: VP from TAS}
    \text{VA}_{\text{TAS}}(\tau) =\bigg|\frac{\int d\omega\ \text{Re}\left[\sigma^{\text{cross}}_{\text{TAS}}(\omega,\tau)-\sigma_{\text{TAS}}^{\text{same}}(\omega,\tau)\right]}{\int d\omega\ \text{Re}\left[\sigma^{\text{cross}}_{\text{TAS}}+\sigma_{\text{TAS}}^{\text{same}}(\omega,\tau)\right]}\bigg|,
\end{equation}
where $\sigma^{\text{same}}$ indicates the TOC obtained by pumping and probing with the same chirality of light as in the left columns of Fig. \ref{fig:tas-four}, and $\sigma^{\text{cross}}$ refers to pumping and probing with opposite circular polarizations, as in the right columns of Fig. \ref{fig:tas-four}.
In Fig. \ref{fig:TAS-depol} we perform a TCD calculation using frozen phonons and compare the measure in Eq. (\ref{eq: VP from TAS}) coming from the TCD directly to the valley polarization calculated via Eq. (\ref{eq: VP from occupation}) using the occupation data from the pump only part of the TAS calculation. It is clear that there is good agreement between these two measures.

Finally we perform a TCD calculation using the TDDFT program Octopus \cite{Tancogne-Dejean2020} in a $30\times30$ supercell with a delay spacing $\Delta\tau$ of $5$ fs. Due to the size of this calculation, 1800 ions in periodic boundary conditions, it was computationally infeasible to do the full calculation with dynamical ions. However we can compare to a full MTEF calculation sampling the $\Gamma$ point optical phonons with ten dynamical calculations in a primitive cell in gray. The fully \textit{ab initio} calculations show a smaller degree of valley polarization at the peak of the pulse, and the supercell calculations show an extremely rapid depolarization, on a timescale roughly two to three times faster than the tight binding model. As expected, the displacements in the primitive cell calculation, while demonstrating some fluctuation of the signal, do not depolarize due to being restricted to $\Gamma$ point phonons, incapable of scattering electronic states from $K^-$ to $K^+$.

\subsection{Tracking Valley Asymmetry with Optical Harmonic Polarimetry}\label{sec: transient optical harmonic polarimetry}
While testing the predictions of the above TCD calculation experimentally is in principle possible -- as the generation of circularly polarized light in the range of the experimentally measured hBN gap of $6$eV can be achieved via high harmonic generation (HHG) \cite{Klemke2019} -- a much easier measure of the valley asymmetry has been proposed by Jim\'{e}nez-Gal\'{a}n and colleagues \cite{Jimenez-Galan20,Mrudul2021} and recently experimentally tested by Mitra et. al \citep{Mitra2023} which is based off the ellipticity of harmonics generated by a linearly polarized, high-intensity, off-resonant probe. In this section we recreate this experiment \textit{in silico} and test the robustness of this measure when including phonon induced scattering in the simulation. 

The basic idea was explained succinctly in the SI of \cite{Jimenez-Galan20}: when driving the system with an off-resonant probe aligned in the $\Gamma-M$ direction of the BZ, the current response induced parallel to the driving field ($j_{\parallel}$) will not be affected by the population of the $K^+/K^-$ valleys, however the current response in the perpendicular direction ($j_{\perp}$) must be. The contribution to $j_{\perp}$ is proportional to the anomalous Hall conductivity arising from conduction band population in regions of non-zero Berry curvature. For equal $K^+$ and $K^-$ valley populations the anomalous Hall current will have exactly counteracting contributions from these two regions. For unequal population, a non-zero $j_{\perp}$ will emerge, which has been used as an experimental measure for valley polarization in TMDs for nearly a decade \cite{Mak2014}. 

Furthermore, since the Berry curvature around the two valleys have opposite sign, the anomalous current arising from a population imbalance around either valley will always be completely out of phase ($\pi$) with respect to one another, and both $\pi/2$ out of phase with respect to $j_{\parallel}$. This means that the radiation emitted as a result of this current will have an elliptical polarization, with ellipticity $\epsilon\in\left[-1,1\right]$ corresponding to the $K^-$/$K^+$ population imbalance, and gives an all-optical measure of the valley asymmetry. Mitra et. al \cite{Mitra2023} use this phenomenon as a measure of the degree of valley asymmetry in hBN following a bichromatic `trefoil' light pulse -- which has been proposed as a tunable driver of valley selective excitation in monolayer hBN and graphene, as well as bulk TMDs 
 and twisted bilayer graphene \cite{Tyulnev2023, Chen2023} -- finding a small but apparently valley selective signal about 100 fs after excitation. 
 
While the ellipticity of emitted harmonics has been proposed to detect other system properties such as topologically insulating phases, the utility of this measure has been called into question by \textit{ab initio} simulations due in part to the many technical difficulties related to the generation and interpretation of HHG signals, even in theoretically ideal conditions \cite{Tancogne-Dejean2017,Tancogne-Dejean2017_2,Freeman2022, Goulielmakis2022,Neufeld2023_2}. Furthermore the HHG spectrum has been found to be very sensitive to $\Gamma$ point phonon distortions  \cite{Neufeld2022}. Given these open questions in the literature, along with the results in Sections \ref{sec: excited carrier relaxation} and \ref{sec: transient circular dichroism} indicating that phonons should induce ultrafast sub-30 fs homogenization of any valley asymmetry in monolayer hBN, in this section we test the robustness of this measure of valley asymmetry upon inclusion of phonon degrees of freedom and the time dependence of the asymmetric valley population signal following a trefoil pump in our TB and TDDFT approaches.

\subsubsection{Robustness of Harmonic Ellipticity under Phonon-Induced Valley Equilibration}
We define the intensity of light emitted due to the non-equilibrium current in a given Cartesian direction $x,y,$ as:
\begin{equation}\label{eq: hhg intensity}
|I_{x,y}(\omega)|^2 = \bigg|\int_{t_0}^{t_f} dt e^{i\omega t}m(t, \frac{1}{2}\left[t_f-t_0\right])\ \partial_t j_{x,y}(t) \bigg|^2,
\end{equation}
where $m(t,x)$ is a mask function. For clean high harmonic generation, we utilize a probe pulse which is far from resonance with the energy gap in hBN, $\omega_{\text{probe}} << E_{g}$, allowing us to drive the system at very high intensities, and utilize a `super-sine' envelope for both the probe pulse and the Fourier transform \cite{Neufeld2019}: 
\begin{equation}\label{eq: super-sine mask}
m(t,\tau) = \left(\sin\left(\pi\frac{t-\tau}{T_{\text{probe}}}\right)\right)^{\left(\frac{\big|\pi\left(\frac{t-\tau}{T_{\text{probe}}}-\frac{1}{2}\right)\big|}{\text{w}}\right)}
\end{equation}
where $\text{w}=0.75$, $\tau$ is the center of the mask, and $m(t,\tau)$ is defined to be zero when $|t-\tau|\geq T_{\text{probe}}/2$. This mask is useful as it begins and ends exactly at $0$ while having a short and smooth ramp time, maximizing the number of optical cycles present at full intensity. We drive the system along the mirror axis, parallel to the B-N bond, and define this to be the $y$ direction corresponding to pumping along the $\Gamma-M$ line in the BZ. The probe is depolyed at multiple delays $\tau$ relative to the center of a pump envelope:
\begin{equation}
\mathbf{A}(t;\tau_{\text{probe}}) = \frac{c\sqrt{I_{\text{probe}}}}{\omega_{\text{probe}}}m(t,\tau)\cos(\omega_{\text{probe}}t + \phi)\hat{\mathbf{y}}
\end{equation}
Where $I_{\text{probe}}$ is the probe intensity, $c$ is the speed of light, and $\phi$ is the CEP, which is always fixed from the start time, i.e., $\phi=\omega_{\text{probe}}(\tau-T_{\text{probe}}/2)$.

The ellipticity of emitted light calculated by Eq. (\ref{eq: hhg intensity}) at a given energy $\omega$ can be determined via the Stokes parameters (\cite{Fleischer2014} Eq. SI.2):
\begin{equation}\label{eq: ellipticity}
\begin{split}
    \epsilon(\omega) &= h\frac{|I_x|^2 + |I_y|^2 - \sqrt{(I_x - I_y)^2 + 4I_xI_y\cos^2\left(\phi_y-\phi_x\right)}}{|I_x|^2 + |I_y|^2 + \sqrt{(I_x - I_y)^2 + 4I_xI_y\cos^2\left(\phi_y-\phi_x\right)}}\\
    \phi_x &= \text{arg}\left(I_x\right),\null\quad\phi_y = \text{arg}\left(I_y\right),\\
    h &= \text{sign}\left[\bigg|I_x+iI_y\bigg| - \bigg|I_x-iI_y\bigg|\right],
\end{split}
\end{equation} 
where $h$ is the helicity of the signal corresponding to the handedness: -1 for right, 1 for left, 0 for linear, and $\phi_{x,y}$ is the phase of the signal. The ellipticity ranges continuously from $[-1,1]$ defining fully right circularly polarized light to fully left circularly polarized light. 

One can define the ellipticity of a given harmonic by taking the normalized harmonic yield for each harmonic $n$, weighted by the ellipticity:
\begin{equation}\label{eq: elliptical yield}
    \text{Elliptical Yield}\ (n) = \frac{\int_{n-1/2}^{n+1/2}d\omega \epsilon(\omega)\left(|I_x|^2 + |I_y|^2\right)}{\int_{n-1/2}^{n+1/2}d\omega\left(|I_x|^2 + |I_y|^2\right)},
\end{equation}
where the integral goes from the energy at  $(n-1/2)\omega_{\text{probe}}$ to $(n+1/2)\omega_{\text{probe}}$. With these definitions in hand we can test the ellipticity of the harmonics generated in hBN after irradiation by the on-resonant pump from sections \ref{sec: excited carrier relaxation} and \ref{sec: transient circular dichroism} which, as already demonstrated, produces very strong valley asymmetry. 

\begin{figure}
    \centering
    \includegraphics[width=\linewidth]{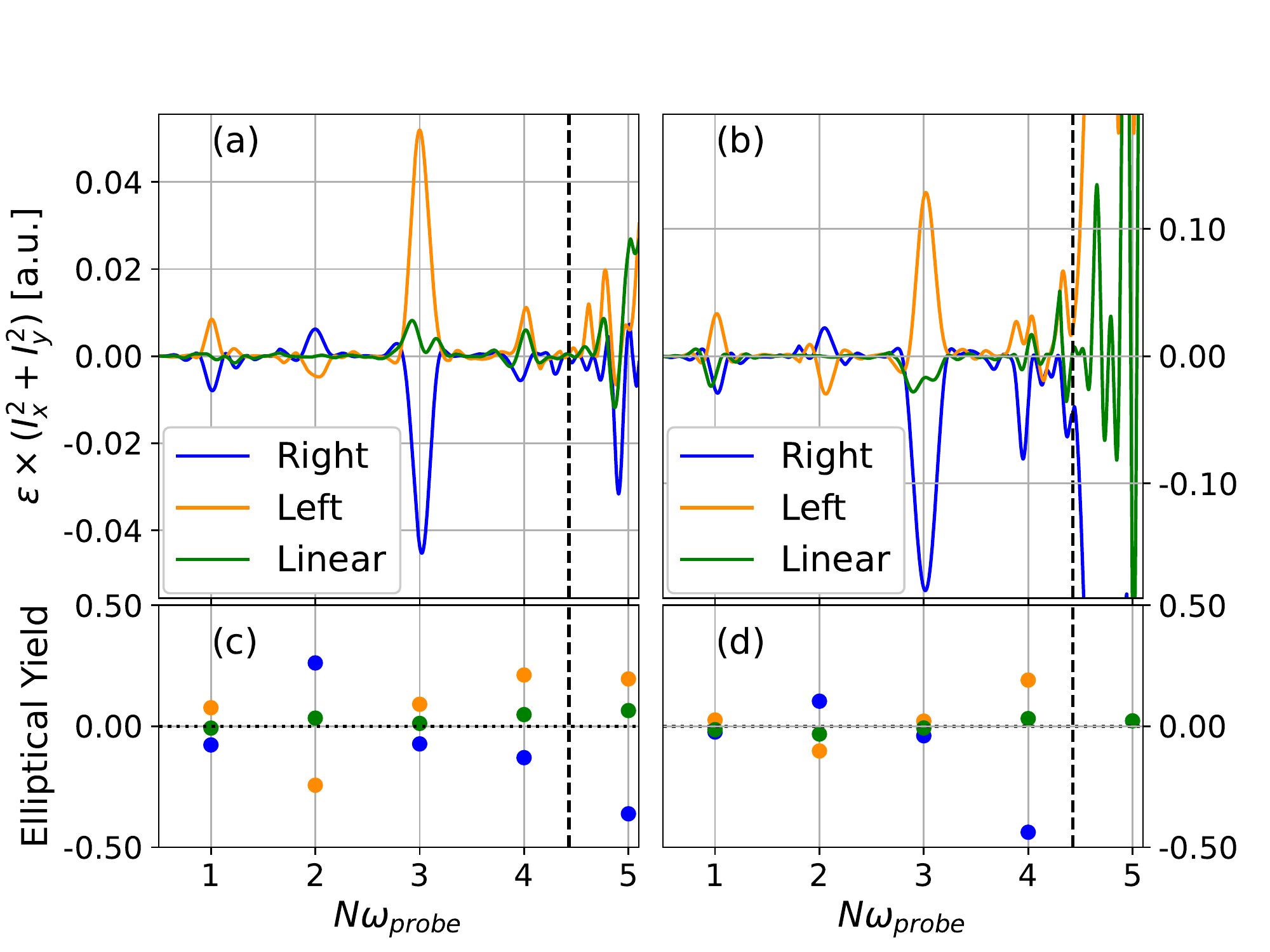}
    \caption{The intensity weighted ellipticity and corresponding elliptical yield of the harmonics generated in a fixed-equilibrium geometry calculation 25 fs after excitation with left circularly, right circularly or linearly polarized light using the same pump parameters as in sections \ref{sec: excited carrier relaxation} and \ref{sec: transient circular dichroism}. The vertical dashed line indicates the conduction band edge at $4.43$ eV. (a): the TB model results, (b): TDDFT results, (c): the elliptical yield of panel (a) calculated with Eq. (\ref{eq: elliptical yield}), (d): the elliptical yield of panel (b). }
    \label{fig:elliptical_yield}
\end{figure}

In Fig. \ref{fig:elliptical_yield}, we see the HHG spectrum for a fixed-equilibrium geometry calculation, 25 fs after being pumped with left circulalry, right circularly or linearly polarized light at $\omega_{\text{pump}}=5$ eV and $T_{\text{pump}}\approx 8.3$ fs, generated by a probe with $\omega_{\text{probe}}=1$ eV, and $T_{\text{probe}}=30$ fs. As one would expect, the fundamental matches the polarization of the driving probe field, while there is flipping of the ellipticity at each subsequent harmonic, until reaching the band edge at $4.43$ eV, whereupon the conduction band electron response dominates the signal. These results hold for both the TB and the TDDFT spectra, though there are naturally some quantitative differences, in particular concerning the intensity of the emitted light. Under linearly polarized pumping, the ellipticity of the spectrum is effectively flat for the TB model, with a small signal at the $3^{\text{rd}}$ harmonic, however the TDDFT results show an elliptical response at both the fundamental and $3^{\text{rd}}$ harmonic, which nonetheless is washed out in the yield calculation. Although the intensity weighted ellipticity can be quite large, due to the normalization in Eq. (\ref{eq: elliptical yield}) the calculated yield does not necessarily reflect this. 

\begin{figure}
    \centering
    \includegraphics[width=\linewidth]{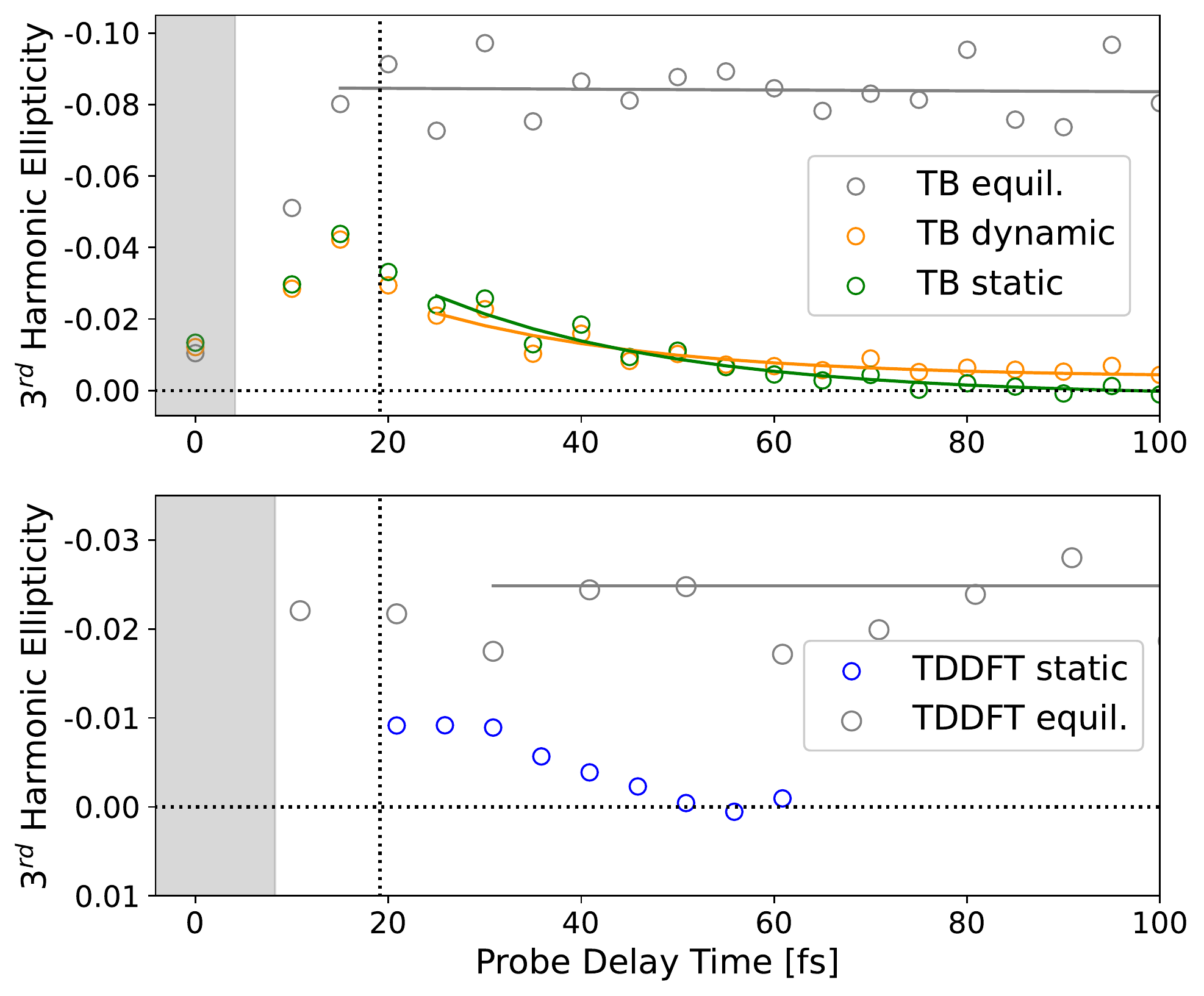}
    \caption{The elliptical yield of the $3^{\text{rd}}$ harmonic calculated with Eq. (\ref{eq: elliptical yield}), following excitation with the same circularly polarized resonant pump from sections \ref{sec: excited carrier relaxation} and \ref{sec: transient circular dichroism}. The grey region corresponds to the duration of the pump, while the vertical dotted line indicates the final time of pump-probe overlap.} 
    \label{fig:HHG-on-resonant}
\end{figure}
Although the $3^{\text{rd}}$ harmonic signal is small, we find that it provides the cleanest time resolved information when scanning over probe delays, shown in Fig. \ref{fig:HHG-on-resonant}. Here we see precisely the same behavior as in the previous results. When the ions are fixed at the equilibrium geometry, there is a fluctuation in the signal for both the TB and TDDFT results, but it remains non-zero due to a lack of valley asymmetry decay channels. In contrast when including phonon dynamics either via MTEF or static disorder there is a suppression of the initial signal followed by a rapid decay corresponding to the equilibration of valley population seen in Fig. \ref{fig:valley-polarization-from-occupation}. 

\subsubsection{Robustness of Trefoil Valley Selectivity under Phonon-Induced Valley Homogenization}
\begin{figure*}
    \centering
    \includegraphics[width=\textwidth]{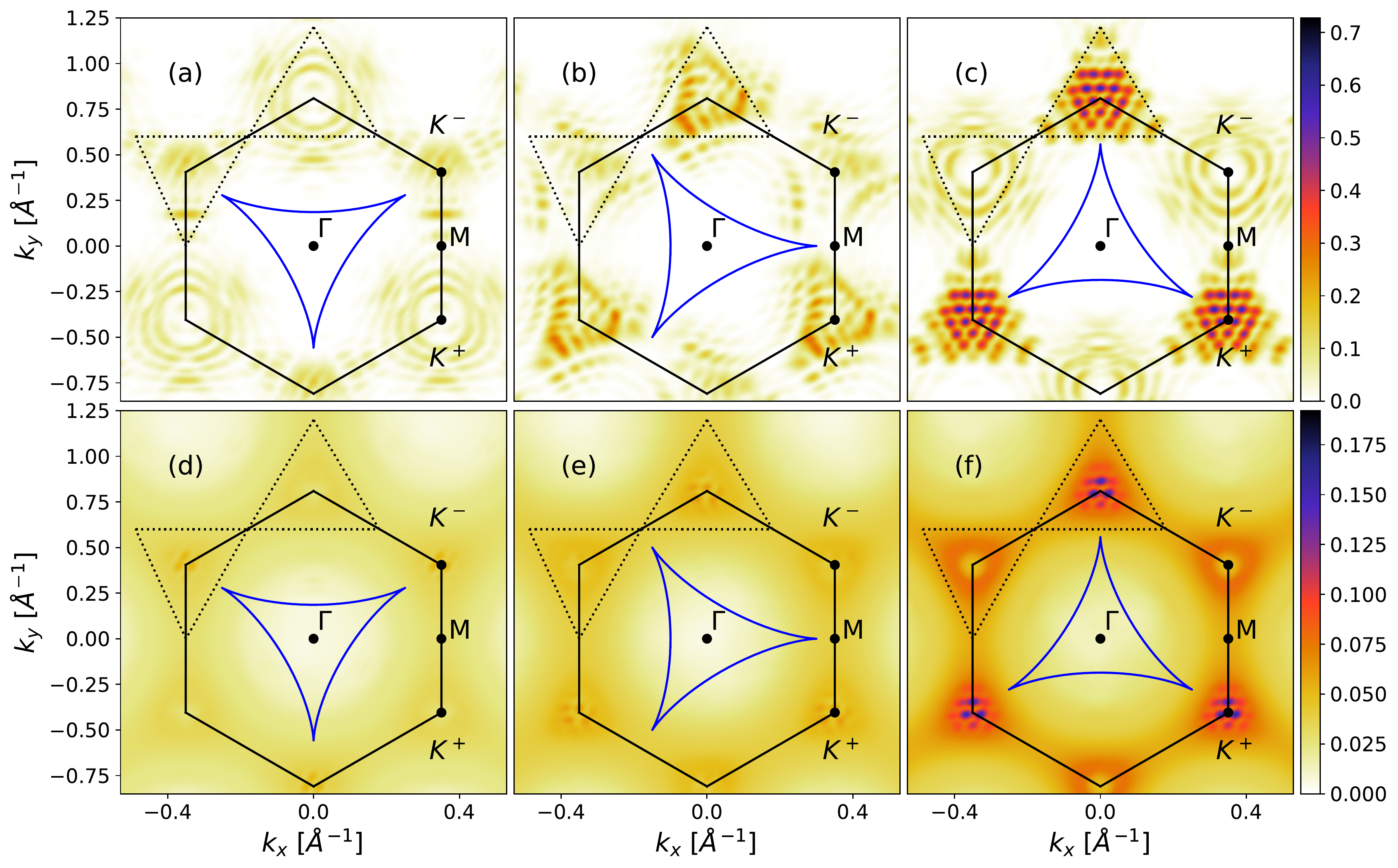}
    \caption{The occupation of the conduction band 50 fs after irradiation with the trefoil pump in the TB model. The path of the trefoil pump in the plane is arbitrarily rescaled and drawn within the BZ as the blue line. Panels (a), (b) and (c) show the results for the static equilibrium geometry when the degree of rotation of the trefoil with respect to the BZ x axis is at $-30^{\circ}$, $0^{\circ}$ and $30^{\circ}$ respectively. Panels (d), (e) and (f) show the MTEF results at the same rotations -- Note the smaller colormap scale. The triangular regions inscribed by the dotted lines are formed by vertices at the high symmetry points between the $K^+/K^-$ points, and are used to define the integration regions for the valley asymmetry calculation in Fig. \ref{fig:Trefoil-decay}.}
    \label{fig:Trefoil-BZ}
\end{figure*}
By combining two counter-rotating circularly polarized fields with a fundamental frequency $\omega_{\text{tr}}$ and it's second harmonic $2\omega_{\text{tr}}$ with a relative amplitude of 2:1, and a delay $t_d$ of the $2\omega_{\text{tr}}$ field relative to the fundamental, one obtains a laser with a triangular `trefoil' shape in the plane which can be arbitrarily rotated by tuning $t_d$. The orientation of the trefoil was shown numerically via the semi-conductor Bloch equation to preferentially excite electrons into the $K^+$ or $K^-$ valleys in monolayer hBN \cite{Jimenez-Galan20, Mitra2023}. The explicit form of the trefoil gauge field we use is:
\begin{equation}
    \begin{split}
        \mathbf{A}_{\text{trefoil}}(t) = A_0 m(t) &\left(\text{Re} \left[\frac{2}{3}e^{-i\omega_{\text{tr}}t} + \frac{1}{3}e^{i2\omega_{\text{tr}}(t-t_{d})}\right]\hat{\mathbf{x}} \right.\\
        &\left.+ \text{Im} \left[\frac{2}{3}e^{-i\omega_{\text{tr}}t} + \frac{1}{3}e^{i2\omega_{\text{tr}}(t-t_{d})}\right]\hat{\mathbf{y}}\right),
    \end{split}
\end{equation}
where again $m(t)$ is an envelope function.  We chose $\omega_{\text{tr}}=0.6$ eV for a duration $T_{\text{pump}}=30$ fs as in the experimental paper, with a super-sine envelope and an intensity of $1.67\times10^{12}$ W/cm$^2$. The effect of driving the TB system with this laser is shown via the conduction band occupations plotted in Fig \ref{fig:Trefoil-BZ}. 

In Fig. \ref{fig:Trefoil-BZ}, panels (a-c) show the results for the static equilibrium geometry with the trefoil pulse oriented at $-30^{\circ}$, $0^{\circ}$, and $30^{\circ}$ rotations relative to the $x$ axis in the BZ. The excitation at $K^+$ for the $30^{\circ}$ rotation in Fig. \ref{fig:Trefoil-BZ}(c) is quite strong and agrees qualitatively with Fig. 2e of Ref. \cite{Jimenez-Galan20}, which shows a similar pumping frequency. At the opposite tuning in panel (a) when $K^-$ should be more populated, there is indeed some excitation, and looking closely one can see that it's texture also resembles the `grape-cluster' structure of the $K^+$ valley in panel (c), although the magnitude of excitation is lower. Qualitatively this reproduces the relative difference in excitation density reported in Fig. 3d and 3e in \cite{Jimenez-Galan20}, confirming that our model also captures this phenomenon. Halfway between these two tunings, in panel (b), one still sees a strong preferential excitement in the $K^+$ valley over the $K^-$ valley.

Incorporating phonon dynamics via MTEF in panels (d), (e), and (f), these trends broadly remain 50 fs after pumping, though the fine structure of the excitation seen away from the $K$ valleys is washed out by electron-phonon scattering, causing a radially symmetric distribution away from the BZ boundaries that decreases towards $\Gamma$. Inclusion of the phonon system via static disorder produces BZ occupations virtually identical to the MTEF results at this timescale. 

\begin{figure}
    \centering
    \includegraphics[width=\linewidth]{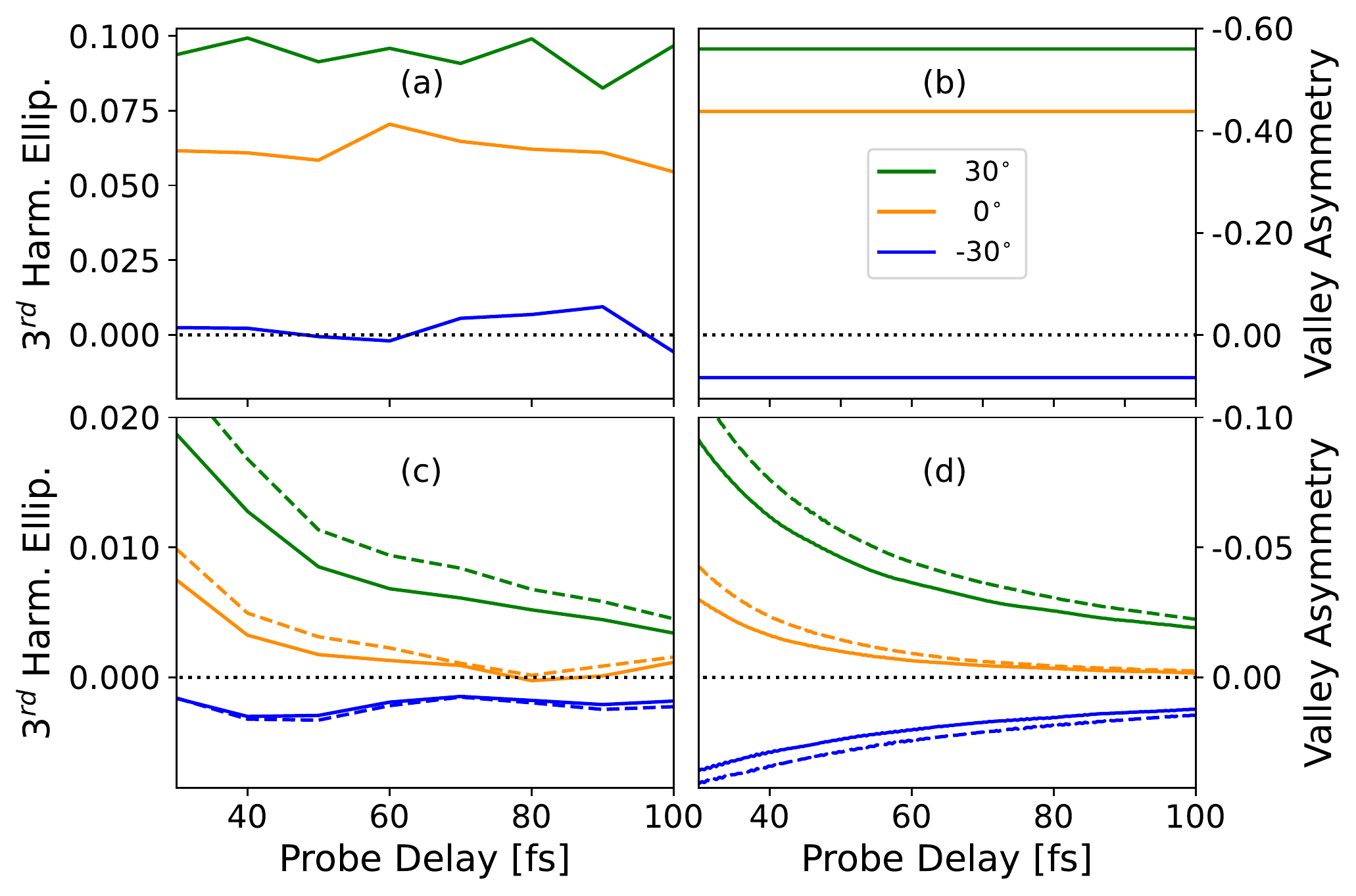}
    \caption{The left column shows the ellipticity yield of the $3^{\text{rd}}$ harmonic generated after irradiation with the three trefoil pulse rotations seen in Fig. \ref{fig:Trefoil-BZ}, compared to the right column showing valley asymmetry calculated by applying Eq. (\ref{eq: VP from occupation}) to the triangular dotted regions in Fig. \ref{fig:Trefoil-BZ}. Panels (a) and (b) show the static equilibrium geometry results while panels (c) and (d) show the MTEF results in bold lines and static results in dashed lines. Note the order of magnitude difference between the y-axes scales for the two approaches.}
    \label{fig:Trefoil-decay}
\end{figure}
In all cases, looking within a small region around $K^{+}/K^-$ when pumping at $-30^{\circ}$ versus $30^{\circ}$ one sees there is indeed an asymmetry in the excitation at the extrema of trefoil orientation. To see how this manifests in the HHG signal, we recreate the experimental setup by utilizing a linearly polarized probe of the same intensity and fundamental frequency as the trefoil pump ($\omega_{\text{probe}} = \omega_{\text{tr}}$) to create an HHG spectrum. As the HHG from the pump alone has no contribution to the $3\omega_{\text{tr}}$ harmonic channel it serves as a natural signal region to study with the probe. In Fig. \ref{fig:Trefoil-decay} we compare the ellipticity  of the $3^{\text{rd}}$ harmonic to the valley asymmetry calculated directly from the conduction band occupation induced exclusively by the trefoil pump for fixed equilibrium geometry in panels (a) and (b), and MTEF/static displacement in panels (c) and (d). Because of the diffuse excitation, we choose integration regions for Eq. (\ref{eq: VP from occupation}) in the BZ formed from vertices at the high symmetry points bisecting the lines between adjacent $K^+/K^-$ valleys, which neatly encapsulates the excitation around $K^+$ for  a $30^{\circ}$ trefoil pump. 

The equilibrium geometry results in panel (b) show that while the valley asymmetry integrated in the conduction band generally shows a tuning of the valley occupations, beginning preferentially in the $K^-$ valley at $-30^{\circ}$ and in the $K^+$ valley at $30^{\circ}$, it is not exact, as at $0^{\circ}$ there should in principle be more equal population, yet there is clearly a preferential excitation.  This is partially reflected in the ellipticity in panel (a), where there is a strong polarization signal at $30^{\circ}$ and $0^{\circ}$, however, the small degree of $K^-$ population under $-30^{\circ}$ pumping fails to contribute to a significant negative elliptical emission. 

These signals become cleaner when looking at the MTEF/static results in panel (c). There is a much smaller degree of polarization at $0^{\circ}$, while $-30^{\circ}$ is negative throughout. In all pump cases, there is a decay in the signal over the $70$ fs depicted which is commensurate with the decay of valley asymmetry seen in panel (d). Comparing the strength of the signal when using oppositely tuned pumps, these simulations indicate that the relative signal difference is very weak after 100 fs. 

\section{Conclusion}
We have derived a general expression for the Wigner transformation of the phonon thermal density matrix applicable to the phonon dispersion for real materials taken from \textit{ab initio} calculations, and utilized it to perform MTEF calculations in both a reciprocal space TB model and a real space TDDFT supercell approach. This methodology allowed us to simulate the relaxation of asymmetrically excited charge carriers in hBN, to track this through the transient absorption spectra using circularly polarized probes, and to recreate an experimental observation of harmonic ellipticity as a measure of valley population imbalance. The inclusion of phonon degrees of freedom, either dynamically (including anharmonic effects) or statically, produced fundamentally different spectroscopic simulation results due to the complete exclusion of a critical decay channel in the common equilibrium geometry framework.

We discussed connections between the MTEF method and static displacement approaches, finding that MTEF is analogous to an extension of the reciprocal space William-Lax / Zacharias-Giustino coordinate distribution to include nuclear velocities/momenta. We have also demonstrated that MTEF and static displacement methods can capture the phonon driven sub-30 fs valley depolarization on time scales commensurate with the TDBE results. Further, we showed that while static displacement methods are restricted to elastic scattering, MTEF  suffers from unrealistic electron heating at long time scales due to ZPE leakage. Despite this, we showed that the temperature dependence of the short time scale relaxation of excited charge carriers agrees well across these simulation methods, exhibiting a plateau in inter-valley relaxation lifetimes at low temperature that agrees relatively well with experimental measurements in similar systems \cite{Zeng2012,Molina-Sanchez2017}. Furthermore we found that these results converge with a very small number of samples.

This work extends MTEF into the domain of the \textit{ab initio} treatment of periodic systems, providing a starting point for the hierarchy of semiclassical dynamics approaches which build on from the mean field limit, correcting for some of it's most serious shortcomings. However, even with issues in the long time-scale behavior of MTEF, this simulation method provides a unique tool to study the ultrafast response behavior of systems with strong electron-phonon coupling in laser driven regimes far from thermal equilibrium. Inclusion of phonon dynamics has the potential to be utilized to incorporate phonon dynamics into fully \textit{ab initio} simulations of parametric driving, Floquet engineering and driven phase transitions, while already inclusion of static disorder can substantially change predictions of spectroscopic measurements via inclusion of elastic electron-phonon scattering under arbitrarily complex pump-probe configurations.

The rapidity with which our results converge for the results presented here is highly encouraging for other far from equilibrium observables, and given the simplicity of incorporating our method into existing real time simulation protocols of non-equilibrium driven phenomena in quantum materials, as well as the fundamental qualitative differences in the resulting temperature dependent dynamics resolved in a systematic framework, we think that this will be a valuable tool going forward. 

\section{Computational Details}
\subsection{Tight Binding Model}
The lattice parameter of $a_0=4.734$ Bohr was obtained from cell relaxation in Quantum Espresso. The phonon frequencies and displacements were calculated on a Monckhorst-Pack (MP) grid of $30\times 30$ using Quantum Espresso, with a $64\times 64$ MP electronic k point grid. We utilize a $30^2\ \mathbf{k}$ and $\mathbf{q}$ grid in the BZ for our Equilibrium, MTEF and static TB simulations in sections \ref{sec: excited carrier relaxation} and \ref{sec: transient circular dichroism} which is sufficiently dense to converge our results compared to a $36^2$ grid. In section \ref{sec: transient optical harmonic polarimetry} we used a $60\times 60$ $\mathbf{k}$-grid. 

The tight binding parameters $t_0$ and $\Delta_{\alpha}=\pm|\Delta|$ were fit from the conduction band calculated over a $30\times 30$ MP grid in the TDDFT code Octopus according to the analytical dispersion relation of the equilibrium geometry tight binding model:
\begin{equation}\label{eq:tight-binding-dispersion}
    E_{\mathbf{k}} = \sqrt{\Delta^2 + t_0^2|\gamma(\mathbf{k})|^2},\null\quad \gamma(\mathbf{k}) = \sum_{\boldsymbol{\delta}}e^{i\mathbf{k}\cdot\boldsymbol{\delta}}.
\end{equation}
An LDA functional was used with Hartwigsen-Goedeker-Hutter (HGH) norm-conserving pseudopotentials with the simulation box having a real space grid spacing of 0.35 Bohr, and being periodic only in two dimensions with a vacuum of $1a_0$ on either side of the monolayer. The band gap of $E_{g} = 2|\Delta|$ corresponds to the uncorrected LDA direct gap of $4.43$ eV, and $t_0=2.68$ eV. Although hBN has rather high energy phonons compared to many other materials, the electronic gap remains an order of magnitude larger, meaning that none of the physics of the electron dynamics within the upper band presented here is affected by not correcting this gap.

The hopping parameter $b$ was calculated by fitting the bands to Eq. (\ref{eq:tight-binding-dispersion}) for a range of ionic configurations with maximum displacement of $3\%$ of the lattice constant, about 0.14 Bohr from equilibrium in steps of 0.005 Bohr, using a first order expansion of Eq.(17). The resulting value of $b=2.87$ is quite close to the value suggested in \cite{Droth2016} ($b = 3.3$), where it was approximated from Slater-Koster parameters between nearest and next nearest neighbors, and is also in a similar range to the value of $3.37$ which has been estimated for graphene \cite{Mohanty2019}. 

The tight binding model is written in Python and C++/CUDA, and is available at gitlab.com under 
\begin{verbatim}
kevin.lively_mpsd/graphene-tight-binding
\end{verbatim}

\subsection{MTEF Simulations}
We integrated the MTEF equations of motion with an fourth order Runge-Kutta algorithm for the electronic portion simultaneously with a velocity-Verlet type scheme for the phonon configuration, using a time step of $0.1$ a.u..

\subsection{TDDFT Simulations}
The TDDFT simulation was done with a time step of $5\times10^{-3}$ fs using a $16^{\text{th}}$ order Lanczos expansion of the exponential propagator in an approximated enforced time reversal symmetry framework. Only the $\Gamma$ point was included in the $30\times30$ supercell BZ, i.e. equivalent to a $30\times30$ $\mathbf{k}$ grid in the primitive cell BZ under the BvK boundary conditions. We used the same grid spacing, pseudopotentials, and simulation box geometry that were used to fit the tight binding model.

\subsection{TDBE Simulations}
All of our TDBE results are generated using input from the tight-binding model Eq. (\ref{eq:Hamiltonian reciprocal space}) in the main text, with band energies $\epsilon_{n\mathbf{k}}=\pm|\epsilon_l(\mathbf{X}=0)|$, and electron-phonon matrix elements taken via projection of the coupling terms onto the band states $g_{mn}^{\nu}(\mathbf{k},\mathbf{q}) = \braket{\psi_{n\mathbf{k}}|l_{\mathbf{q}\nu}\hat{M}(\mathbf{q}\nu)|\psi_{m\mathbf{k}}}$. 

The delta functions used in the TDBE scattering rate equations are approximated by Gaussian distributions $\delta(x)\approx \frac{1}{\eta\sqrt{\pi}}e^{-\left(\frac{x}{\eta}\right)^2}$ with $\eta=10$ meV, and the resultant scattering rates converged with respect to number of $\mathbf{k}/\mathbf{q}$ points and $\eta$ \cite{Zhou2016,Tong2021}. The equations of motion themselves are integrated using a fourth order Runge-Kutta algorithm with $f_{n\mathbf{k}}(t),\ n_{\mathbf{q}\nu}(t)$ inserted into each derivative and a timestep of $1$fs. The initial excited electronic carrier population is set via interpolation of the carrier occupations from the equilibrium geometry using a $72^2$ $\mathbf{k}$ MP grid just after the cessation of the pulse onto a denser $90^2$ $\mathbf{k}$ grid. Because we are working with highly non-thermal electronic distributions in a system with extremely strong electron-phonon coupling on top of this gaussian approximation, we find there is a slight drift in the total excited population, and that the ratio of population drift to total population decreases as the initial total excited population decreases. Therefore with the exception of Fig. \ref{fig:occupation-4panel}(d), which is included on the same scale for illustrative purposes, the initial TDBE electronic population distribution is set as $1/100^{\text{th}}$ of the MTEF occupation at the end of the pump, i.e. $f_{c\mathbf{k}}^{\text{TDBE}}(T_{\text{pump}})=0.01f_{c\mathbf{k}}^{\text{MTEF}}(T_{\text{pump}})$.  

\subsection{Transient Optical Conductivity}\label{section: MTEF-TAS}
The simulation protocol for calculating the transient optical conductivity is as follows: \begin{enumerate}
    \item Sample a phonon system initial condition and initialize the electronic system according to Eq. (\ref{eq:electronic-init}), giving a total system initial condition IC$_0$. 
    \item Expose the system to a probe of a given chirality $s_{\text{probe}}$ from IC$_0$, and propagate for a duration of $T_f$.
    \item Reset to IC$_0$ and propagate the system under the influence of the pump to the maximum delay time desired $\tau_{\text{max}}+T_{f}$. At each time $\tau$ that one wants to have data for, save the instantaneous state of the system IC$_{\tau}$. 
    \item Load each IC$_{\tau}$ and propagate under the same pump, but with an added probe of chirality $s_{\text{probe}}$ centered at $\tau+T_{\text{probe}}/2$.
    \item Apply Eqs. (\ref{eq: TAS current}-\ref{eq: final TOC}), using the response from step (2) to calculate $\sigma_{\text{equil}}(\omega)$. 
    \item Repeat steps (2-5) with a probe of the opposite chirality. 
    \item Repeat steps (1-6) for $N_t$ samples and average the results
\end{enumerate}

\subsection{Graphical Representation of Data}
The data in figures \ref{fig:occupation-4panel}, \ref{fig:tas-four} and \ref{fig:Trefoil-BZ} were interpolated with the bicubic interpolation function of matplotlib.

\section*{Acknowledgements}
The authors would like to acknowledge Ofer Neufeld for helpful discussions on calculating the ellipticity of HHG signals, Jonathan Mannouch for helpful discussions about semi-classical dynamics, and Andrey Geondzhian for valuable discussions. We acknowledge financial support from the Cluster of Excellence `CUI: Advanced Imaging of Matter’- EXC 2056 - project ID 390715994, SFB-925 "Light induced dynamics and control of correlated quantum systems" – project 170620586  of the Deutsche Forschungsgemeinschaft (DFG) and Grupos Consolidados (IT1453-22). We acknowledge support from the Max Planck-New York City Center for Non-Equilibrium Quantum Phenomena. The Flatiron Institute is a division of the Simons Foundation.
\bibliography{bibliography}

\newpage
\setcounter{section}{0}
\setcounter{equation}{0}
\setcounter{figure}{0}

\begin{widetext}
\renewcommand{\theequation}{SI.\arabic{equation}}
\renewcommand{\thefigure}{SI.\arabic{figure}}    
\renewcommand{\thesection}{SI.\arabic{section}}    
\renewcommand{\thetable}{SI.\arabic{table}}  
\begin{center}
\textbf{Supplemental Information: Revealing Ultrafast Phonon Mediated Inter-Valley Scattering through Transient Absorption and High Harmonic Generation Spectroscopies}
\end{center}  

\section{Details of the Phonon normal mode description}
We follow primarily the convention from Feliciano Giustino's  Rev. Mod. Phys. \cite{Giustino2017}, restating many definitions here to make the text self-contained. These definitions are used in the derivation of the phonon wigner distribution and the electron-phonon coupling matrix elements.
\subsection{The Born-Von K\'{a}rm\'{a}n Boundary Conditions}\label{section: SI BvK}
The crystalline primitive unit cell is defined by the primitive lattice vectors $\mathbf{a}_i$, for $i=\{1,\ldots,l\}$ for the number of periodic dimensions $l$ and the $p^{\text{th}}$ unit cell is specified by $\mathbf{R}_p = \sum_i n_i\mathbf{a}_i$ with integers $n_i\in [0,N_i-1]$. The BvK supercell contains $N_p=\prod_iN_i$ primitive unit cells. The primitive vectors of the reciprocal lattice are denoted by $\mathbf{b}_j$, fulfilling the duality condition $\mathbf{a}_i \cdot\mathbf{b}_j=2\pi\delta_{ij}$. Consider Bloch wave vectors $\mathbf{q}_i$ defined on a uniform grid in the first Brillouin zone, $\mathbf{q}=\sum_i (m_i/N_i)\mathbf{b}_i$ with integers $m_i\in [0,N_i-1]$. From these definitions we have the following sum rules:
\begin{equation}\label{eq:BvK sum rules}
\begin{split}
    \sum_q \exp(i\mathbf{q}\cdot\mathbf{R}_p) &= N_p\delta_{p,0}\\
    \sum_p\exp (i\mathbf{q}\cdot\mathbf{R}_p) &= N_p\delta_{q,0}.
\end{split}
\end{equation}

\subsection{Normal Mode Coordinates}\label{section: SI Normal Mode Description}
Within the BvK boundary conditions atoms are identified by their positions w/rt the primitive cell, $\mathbf{R}_{\alpha p}^0 = \mathbf{R}_p + \mathbf{R}_{\alpha}^0$, where $p = 1,\ldots, N_c$ identifies the primitive cell, $\alpha$ indicates the specific atom within the primitive cell, $N_c$ is the number of atoms within each primitive unit cell, and $\mathbf{R}_{\alpha}^0$ denotes the equilibrium position within the primitive unit cell defined by being a minimum energy configuration for a given lattice configuration. We can further identify small displacements from these positions via $\delta\mathbf{R}_{\alpha p} = \mathbf{R}_{\alpha p}-\mathbf{R}_{\alpha p}^0$. For small displacements from the minimum energy configuration, we can write the potential as 
\begin{equation}\label{eq:BvK boundary conditions}
\begin{split}
    U &= U_0 + \frac{1}{2}\sum_{\alpha p,\alpha 'p'} \frac{\partial^2 U}{\partial\mathbf{R}_{\alpha p}\partial\mathbf{R}_{\alpha 'p'}}\bigg\rvert_{\mathbf{R}^0_{\alpha,p},\mathbf{R}^0_{\alpha',p'}}\delta\mathbf{R}_{\alpha p}\delta\mathbf{R}_{\alpha 'p'}\\
    &=U_0 + \frac{1}{2}C_{\alpha p,\alpha 'p'}\delta\mathbf{R}_{\alpha p}\delta\mathbf{R}_{\alpha 'p'},
\end{split}
\end{equation}
where we have defined the so called Interatomic Force Constant matrix (IFC) as $\doubleunderline{C}$, with $C_{\alpha p,\alpha 'p'}\in\mathbb{R}^{d\times d}$. Treating the nuclear positions as operators, we define canonical momenta $\mathbf{P}_{\alpha p}$ by the canonical commutation relation $[\mathbf{R}_{\alpha p},\mathbf{P}_{\alpha p}] = i\hbar\delta_{\alpha ,\alpha '}\delta_{p,p'}$ and write the Hamiltonian operator for the nuclei in real space as
\begin{equation}\label{eq:phonon hamiltonian}
    \hat{H}_{ph} = \frac{1}{2}\sum_{\alpha p,\alpha 'p'}C_{\alpha p,\alpha 'p'}\delta\mathbf{R}_{\alpha p}\delta\mathbf{R}_{\alpha 'p'} + \sum_{\alpha p}\frac{1}{2M_{\alpha}}\mathbf{P}_{\alpha p}^2
\end{equation} 

Since the IFC must be invariant under any operation which maps between supercells, it obeys certain symmetry operations. We can encode this via the Fourier transform of the IFC, defined as the dynamical matrix \cite{Maradudin1968}
\begin{equation}\label{eq:DynamicalMatrix}
    D_{\alpha,\alpha '}(\mathbf{q}) = (M_{\alpha}M_{\alpha'})^{-1/2}\sum_p C_{\alpha 0,\alpha 'p}\exp(i\mathbf{q}\cdot\mathbf{R}_p),
\end{equation}
where $M_{\alpha}$ is the mass of the $\alpha^{\text{th}}$ ion. The dynamical matrix is hermitian and positive definite allowing real eigenvalues, denoted as $\omega_{\mathbf{q}\nu}^2$
\begin{equation}\label{eq:normalModes}
    \sum_{\alpha'}D_{\alpha,\alpha '}(\mathbf{q})\mathbf{e}_{\alpha'\nu}(\mathbf{q}) = \omega_{\mathbf{q}\nu}^2\mathbf{e}_{\alpha\nu}(\mathbf{q}).
\end{equation}
In classical mechanics, $\mathbf{e}_{\alpha\nu}(\mathbf{q})\in\mathbb{C}^{d}$ correspond to the normal modes of the system, i.e. independent oscillators with characteristic angular frequency $\omega_{\mathbf{q}\nu}$ for each branch $\nu$ and unique primitive cell atom $\alpha$. The eigenvectors and values of the dynamical matrix have the following properties at each $\mathbf{q}$:
\begin{equation}\label{eq:normal mode rules}
    \begin{split}
        \sum_{\nu} e^*_{\alpha'\nu,i}(\mathbf{q})e_{\alpha\nu,j}(\mathbf{q}) &= \delta_{\alpha,\alpha '}\delta_{i,j}\quad \text{(Completeness)}\\
        \sum_\alpha e^*_{\alpha\nu,i}(\mathbf{q})e_{\alpha\nu',j}(\mathbf{q}) &= \delta_{\nu\nu'}\delta_{i,j}\quad\text{(Orthonormality)}\\
        \omega^2_{-\mathbf{q},\nu} &=\omega^2_{\mathbf{q}\nu}\\
        \mathbf{e}_{\alpha\nu}(-\mathbf{q}) &= \mathbf{e}^*_{\alpha\nu}(\mathbf{q}). 
    \end{split}
\end{equation}

By inserting the decomposition of real space displacement into phonon coordinates, Eq. (\ref{eq:complex normal coordinate inverse}), into (\ref{eq:phonon hamiltonian}), and using equations (\ref{eq:BvK sum rules}) and (\ref{eq:DynamicalMatrix}-\ref{eq:normal mode rules}) we obtain the reciprocal space Phonon Hamiltonian in reduced coordinates, Eq. (\ref{eq:phonon hamiltonian real normal coordinates}). We can further rewrite the phonon hamiltonian by introducing the following ladder operators:
\begin{equation}\label{eq:phonon ladder operators}
\begin{split}
    \hat{a}^{x,\dagger}_{\mathbf{q}\nu} &= \frac{1}{\sqrt{2}}\left(\tilde{x}_{\mathbf{q}\nu} - i\tilde{r}_{\mathbf{q}\nu}\right),\null\quad
    \hat{a}^{x}_{\mathbf{q}\nu} = \frac{1}{\sqrt{2}}\left(\tilde{x}_{\mathbf{q}\nu} + i\tilde{r}_{\mathbf{q}\nu}\right)\\
    \hat{a}^{y,\dagger}_{\mathbf{q}\nu} &= -\frac{1}{\sqrt{2}}\left(i\tilde{y}_{\mathbf{q}\nu} +\tilde{s}_{\mathbf{q}\nu}\right),\null\quad
    \hat{a}^{y}_{\mathbf{q}\nu} = \frac{1}{\sqrt{2}}\left(i\tilde{y}_{\mathbf{q}\nu} - \tilde{s}_{\mathbf{q}\nu}\right),
\end{split}
\end{equation}
whose definitions and properties as ladder operators follow from the phonon momentum inversion properties and canonical commutation relations, Eq.  (\ref{eq:normal coordinate inversion properties}) and  (\ref{eq:real normal coordinate commutation relations}). From here it is trivial to rewrite the reciprocal space phonon hamiltonian Eq. (\ref{eq:phonon hamiltonian real normal coordinates}) as
\begin{equation}\label{eq:phonon hamiltonian ladder operators}
    \hat{H}_{\text{ph}} = \sum_{\mathbf{q}\in \mathcal{A},\nu}\omega_{\mathbf{q}\nu}\left(\hat{a}^{x,\dagger}_{\mathbf{q}\nu}\hat{a}^{x}_{\mathbf{q}\nu} + \frac{1}{2}\right)
    + \sum_{\mathbf{q}\in \mathcal{B},\nu}\omega_{\mathbf{q}\nu}\left(\hat{a}^{x,\dagger}_{\mathbf{q}\nu}\hat{a}^{x}_{\mathbf{q}\nu} + \hat{a}^{y,\dagger}_{\mathbf{q}\nu}\hat{a}^{y}_{\mathbf{q}\nu} + 1\right).
\end{equation}

\section{Derivation of the tight-binding electron-phonon coupling term}\label{section: SI EPh-derivation}
Generally the electron-phonon coupling constants $g_{mn}^{\nu}(\mathbf{q})$ are most often incorporated into semi-classical Boltzmann style equations, meaning that only their absolute value is calculated using DFPT \cite{Yan2009}, or only described analytically in small regions around high symmetry BZ points \cite{Malic2011,Winzer2013,Zhang2015,Rana2009}. However, we require the complex value of the coupling throughout the entire BZ in order to have coherent electronic evolution. Therefore we describe our derivation for this in detail.

Starting with the following Hamiltonain:
\begin{equation}
    \hat{H}_W(\mathbf{X}) = H_{\text{ph}}(\mathbf{X})-\sum_{p\delta}t(\mathbf{R}_{\text{b}},\mathbf{R}_{\text{a}})\left(\hat{a}^{\dag}_p\hat{b}_{p+\delta} + c.c.\right) + \sum_{p\alpha}\Delta_{\alpha}\hat{\alpha}^{\dagger}_p\hat{\alpha}_p,
\end{equation}
we expand the exponential dependence of the hopping term on the nuclear coordinates, $t(\mathbf{R}_{\text{b}},\mathbf{R}_{\text{a}})=t_0\exp\left(-b\left[\frac{|\mathbf{R}_{\text{b}}-\mathbf{R}_{\text{a}}|}{d_0} - 1\right]\right)$, to first order, giving us:
\begin{equation}
\hat{H}_W(\mathbf{X}) = H_{\text{ph}}(\mathbf{X})
-t_0\sum_{p\delta}\left(\hat{a}^{\dag}_p\hat{b}_{p+\delta} + c.c.\right) + \sum_{p\alpha}\Delta_{\alpha}\hat{\alpha}^{\dagger}_p\hat{\alpha}_p 
- \frac{t_0b}{d_0}\sum_{p\delta}\hat{\boldsymbol{\delta}}\cdot\left(\delta\mathbf{R}_{p+\delta ,b}-\delta\mathbf{R}_{pa}\right)\left(\hat{a}^{\dag}_p\hat{b}_{p+\delta} + c.c.\right),
\end{equation}
where $\hat{\boldsymbol{\delta}}$ is the unit vector connecting the $a$ sublattice sites to the $b$ sublattice sites. We can rewrite the real space electron-phonon coupling (EPC) hamiltonian as:
\begin{equation}\label{real space eph hamiltonian}
\hat{H}_{\text{e-ph}} = -\frac{t_0b}{a_0}
\sum_{p}\sum_{p'\in\{p,p\pm 1\}} 
\hat{\boldsymbol{\delta}}^0_{p'}\cdot\left(\mathbf{\delta R}_{bp'}-\mathbf{\delta R}_{ap}\right) \left(\hat{a}_p^{\dag}\hat{b}_{p'}+c.c.\right),
\end{equation}
where $\hat{a}_p$ we have organized the supercell labeling such that the nearest neighbors for each primitive cell are labeled as belonging to $p-1,p,p+1$. Furthermore, defining the equilibrium nearest neighbor distance as $\mathbf{R}_{b p}^0 - \mathbf{R}_{ap}^0 = \boldsymbol{\delta}^0$ we can easily write:
\begin{equation}
\begin{split}
\boldsymbol{\delta}^0_{p-1} &= \mathbf{R}_{b, p-1}^0 - \mathbf{R}_{a p}^0 = \boldsymbol{\delta}^0 - \mathbf{a}_2\\
\boldsymbol{\delta}^0_{p} &= \boldsymbol{\delta}^0\\
\boldsymbol{\delta}^0_{p+1} &=  \boldsymbol{\delta}^0 - \mathbf{a}_2 - \mathbf{a}_1,
\end{split}
\end{equation}
with corresponding unit vectors $\hat{\boldsymbol{\delta}}_p^0$. We take the Fourier transform by replacing the real space lattice site operators with their planewave counterparts:
\begin{equation}\label{eq:planewave operator}
\begin{split}
\hat{\alpha}_p &= N_p^{-1/2}\sum_{\mathbf{k}}e^{i\mathbf{k}\cdot \mathbf{R}_{\alpha p}^0}\hat{\alpha}_{\mathbf{k}}
\end{split}
\end{equation}
Note that we make the distinction between the primitive cell lattice sites $\mathbf{R}_p$ and sublattice sites $\mathbf{R}_{\alpha p}^0$ belonging to the $\alpha\in\{A,B\}$ sublattices:  $\mathbf{R}_{\alpha p}^0$. Although we have ionic displacements $\mathbf{\delta R}_{\alpha p}$, these are fundamentally defined w/rt to the sublattice sites $\mathbf{R}_{\alpha p}^0$ which define the periodicity of the crystal, and therefore how the Fourier transform is defined.

Inserting equations (\ref{eq:planewave operator}) and (\ref{eq:complex normal coordinate inverse}) into equation (\ref{real space eph hamiltonian}), we have the following:
\begin{equation}
\begin{split}
\hat{H}_{\text{e-ph}} &= -\frac{t_0b}{d_0}N_p^{-3/2}\sum_{p\delta}\sum_{q\nu}\sum_{\mathbf{k}\mathbf{k}'}\boldsymbol{\hat{\delta}}\cdot\left(\left(\frac{M_0}{M_b}\right)^{\frac{1}{2}}e^{i\mathbf{q}\cdot\boldsymbol{\mu}_{\delta}}\mathbf{e}_{b\nu}(\mathbf{q}) - \left(\frac{M_0}{M_a}\right)^{\frac{1}{2}}\mathbf{e}_{a\nu}(\mathbf{q})\right)z_{\mathbf{q}\nu}\\
&\times\left(e^{i\mathbf{R}_p\cdot\left(\mathbf{q}+\mathbf{k}'-\mathbf{k}\right)}e^{i\mathbf{k}'\cdot\boldsymbol{\delta}}\hat{a}_{\mathbf{k}'}^{\dagger}\hat{b}_{\mathbf{k}} + e^{i\mathbf{R}_p\cdot\left(\mathbf{q}-\mathbf{k}'+\mathbf{k}\right)}\hat{b}_{\mathbf{k}'}^{\dagger}\hat{a}_{\mathbf{k}}\right)\\
&=\sum_{\mathbf{q}\nu\delta}\sum_{\mathbf{k}\mathbf{k}'}g_{\nu}^{\delta}(\mathbf{q})z_{\mathbf{q}\nu}\left(\delta_{\mathbf{q}+\mathbf{k}',\mathbf{k}}e^{i\mathbf{k}'\cdot\boldsymbol{\delta}}\hat{a}_{\mathbf{k}}^{\dagger}\hat{b}_{\mathbf{k}} + \delta_{\mathbf{q}-\mathbf{k}',\mathbf{k}}e^{-i\mathbf{k}'\cdot\boldsymbol{\delta}}\hat{b}_{\mathbf{k}'}^{\dagger}\hat{a}_{\mathbf{k}}\right),
\end{split}
\end{equation}
Where $\mu_{\delta} = \{\mathbf{0}, -\mathbf{a}_2, -\mathbf{a}_2-\mathbf{a}_1\}$ are the primitive lattice vectors connecting primitive cell $\mathbf{R}_p$ to $p'=\{p-1,p,p+1\}$, and in the second line we have summed through $p$, applying the BvK boundary conditions Eq. (\ref{eq:BvK boundary conditions}), and defined the electron-phonon coupling term as
\begin{equation}
    g_{\nu}^{\delta}(\mathbf{q}) \coloneqq 
    -\frac{t_0b}{N_p^{1/2}d_0}\boldsymbol{\hat{\delta}}\cdot\left(\left(\frac{M_0}{M_b}\right)^{\frac{1}{2}}e^{i\mathbf{q}\cdot\boldsymbol{\mu}_{\delta}}\mathbf{e}_{b\nu}(\mathbf{q}) - \left(\frac{M_0}{M_a}\right)^{\frac{1}{2}}\mathbf{e}_{a\nu}(\mathbf{q})\right).
\end{equation}
We can group $g_{\nu}^{\delta}(\mathbf{q})$ and everything in parentheses together and call it the electron-phonon coupling operator $\hat{M}(\mathbf{q},\nu)$.

By summing $\mathbf{q}$ through $\mathcal{C}$, we obtain the actually implemented form of the tight binding Hamiltonian in Eq. (\ref{eq:Hamiltonian reciprocal space}): 
\begin{equation}
\begin{split}
\hat{H}_W(\mathbf{X}) &=  H_{\text{ph}}(\mathbf{X}) + \hat{H}_e +  \sum_{\nu,\mathbf{q}\in\mathcal{A}} x_{\mathbf{q}\nu}\hat{M}(\mathbf{q},\nu) 
	     + \sum_{\nu,\mathbf{q}\in\mathcal{B}} x_{\mathbf{q}\nu}\left(\hat{M}(\mathbf{q},\nu) + \hat{M}^{\dag}(\mathbf{q},\nu)\right) + iy_{\mathbf{q}\nu}\left(\hat{M}(\mathbf{q},\nu) - \hat{M}^{\dag}(\mathbf{q},\nu)\right),
\end{split}
\end{equation}
where $H_e$ gathers the bare hopping and onsite energy terms. This means that the equations of motion for each $x_{\mathbf{q}\nu}^i,\ y_{\mathbf{q}\nu}^i$ trajectory are explicitly
\begin{equation}
\begin{split}
\dot{r}_{\mathbf{q}\nu} = -\bigg\langle\frac{\partial H}{\partial x_{\mathbf{q}\nu}^i}\bigg\rangle &= -\text{Tr}\left[\hat{\rho}_i(t)\left(\hat{M}(\mathbf{q},\nu) + \hat{M}^{\dag}(\mathbf{q},\nu)\right)\right] - \frac{\omega_{\mathbf{q}\nu}}{l_{\mathbf{q}\nu}}x_{\mathbf{q}\nu}\\
\dot{s}_{\mathbf{q}\nu} = -\bigg\langle\frac{\partial H}{\partial y_{\mathbf{q}\nu}^i}\bigg\rangle &= -i\text{Tr}\left[\hat{\rho}_i(t)\left(\hat{M}(\mathbf{q},\nu) - \hat{M}^{\dag}(\mathbf{q},\nu)\right)\right]- \frac{\omega_{\mathbf{q}\nu}}{l_{\mathbf{q}\nu}}y_{\mathbf{q}\nu}.
\end{split}
\end{equation}
Looking at the structure of $\hat{M}(\mathbf{q},\nu)$ one can see that $\hat{M}(\mathbf{q},\nu) + \hat{M}^{\dag}(\mathbf{q},\nu)$ will be hermitian while $\hat{M}(\mathbf{q},\nu) - \hat{M}^{\dag}(\mathbf{q},\nu)$ will be anti-hermitian. Therefore expectation values of the former will always be real and imaginary for the latter.

\section{Electronic Current Operator}\label{section: SI Current operator}
The current density operator can be defined as
\begin{equation}
\hat{\mathbf{j}}_W = \frac{i}{\Omega}\left[\hat{\mathbf{r}},\hat{H}_W(t)\right]
\end{equation}
for the super cell surface area $\Omega$ and the electron position operator
\begin{equation}
\hat{\mathbf{r}} = \sum_{i\in N_p}\sum_{\delta}
\mathbf{r}_i\hat{a}_i^{\dag}\hat{a}_i 
+ \left(\mathbf{r}_i+\boldsymbol{\delta}\right)\hat{b}^{\dag}_{i+\delta}\hat{b}_{i+\delta}
= \sum_{i\in N_p}\sum_{\delta}\mathbf{r}_i\hat{n}_i + \left(\mathbf{r}_i+\boldsymbol{\delta}\right)\hat{n}_{i+\delta}
\end{equation}
Where i is summed over primitive cells and the number operator $\hat{n}_{i+\delta}$ is understood to be a $b$ sublattice number operator to reduce notational clutter. 
We can utilize the definition of the sublattice anticommutation,
\begin{equation}
\begin{split}
\{\hat{\alpha}_i,\hat{\beta}_j^{\dag}\} &= \delta_{i,j}\delta_{\alpha,\beta}\\
\{\hat{\alpha}_i,\hat{\beta}_j\} &= \{\hat{\alpha}_i^{\dag},\hat{\beta}_j^{\dag}\} = 0,\\
\end{split}
\end{equation}
where $\hat{\alpha},\hat{\beta}$ can be either $\hat{a}$ or $\hat{b}$ site operators to derive the identity:
\begin{equation}\label{identity}
\left[\hat{n}_{i},\hat{a}^{\dag}_j\hat{b}_{j+\delta}\right] = \left(\delta_{ij}-\delta_{i,j+\delta}\right)\hat{a}_j^{\dag}\hat{b}_{j+\delta}.
\end{equation}
Taking the commutator and utilizing the identity (\ref{identity}) we obtain
\begin{equation}
\hat{\mathbf{j}}= e\frac{i}{\Omega} \sum_{i\in N_p}\sum_{\delta}\boldsymbol{\delta}\left(t_0 + \frac{t_0b}{a_0}\hat{\boldsymbol{\delta}}\cdot\left(\delta\mathbf{r}_{i+\delta}-\delta\mathbf{r}_i\right)\right)
\left(\hat{a}^{\dag}_i\hat{b}_{i+\delta} - c.c.\right).
\end{equation}
Expanding in plane waves and phonons as before we have:
\begin{equation}
\begin{split}
\hat{\mathbf{j}} &= e\frac{i}{\Omega} \left[t_0\sum_{\mathbf{k}}\sum_{\delta}\boldsymbol{\delta}\left(\hat{a}^{\dag}_{\mathbf{k}}\hat{b}_{\mathbf{k}}e^{i\mathbf{k}\cdot\boldsymbol{\delta}} - c.c.\right)
 + \sum_{\mathbf{q}\nu}z_{\mathbf{q}\nu}\hat{D}(\mathbf{q},\nu)\right]\\
 \hat{D}(\mathbf{q},\nu) &= \sum_{\mathbf{k}}\sum_{\delta}\boldsymbol{\delta}
g_{\nu}^{\delta}(\mathbf{q})
\left(\hat{a}_{\mathbf{k}+\mathbf{q}}^{\dag}\hat{b}_{\mathbf{k}}e^{i\mathbf{k}\cdot\boldsymbol{\delta}} - \hat{b}_{\mathbf{k}}^{\dag}\hat{a}_{\mathbf{k}-\mathbf{q}}e^{-i\mathbf{k}\cdot\boldsymbol{\delta}}\right).
\end{split} 
\end{equation}
We take $\Omega$ to be $N_p$ times primitive cell area $\Omega_0 = 19.5$bohr$^2$

\section{Electronic State Renormalization in Graphene}\label{section: SI graphene}
\begin{figure}
    \centering
    \includegraphics[width=0.8\textwidth]{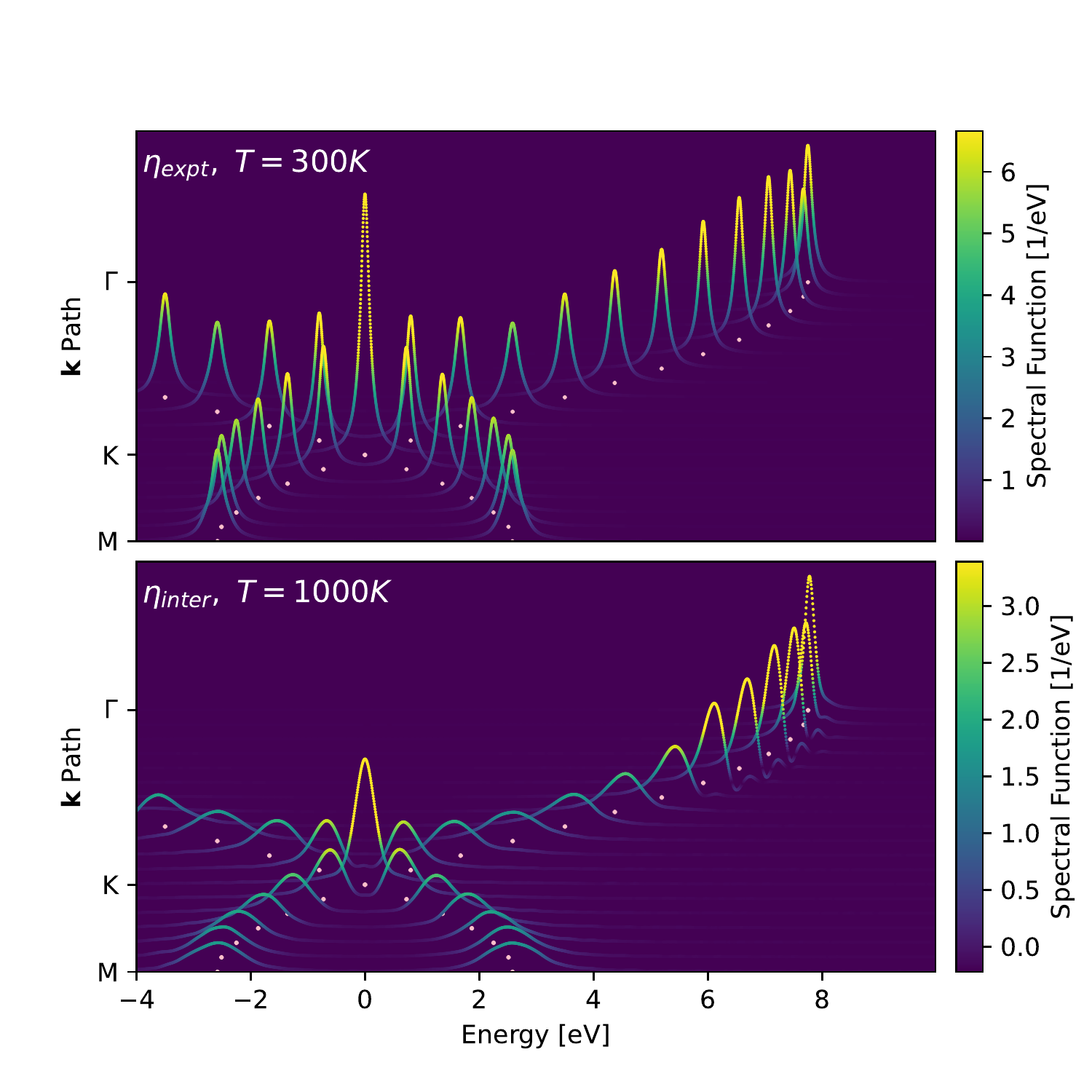}
    \caption{The spectral function $A_{\mathbf{k}}(\omega)$ of graphene calculated in the tight binding model through MTEF, with electron-phonon coupling constants $\eta$ and temperatures matching the simulations from Ref. \cite{Nery2022}, Figure 8. See text for details. }
    \label{fig:Mauri}
\end{figure}
 To provide a sanity check of our derivation of $g_{\nu}^{\delta}(\mathbf{q})$ in the tight binding model, we first apply our tight binding Hamiltonian in Eq. (\ref{eq:Hamiltonian reciprocal space}) to graphene and compare with recent results from Nery and Mauri \cite{Nery2022} using static displacement averaging to calculate the electron-phonon interaction renormalized spectral function $A_{\mathbf{k}}(\omega)$.

Reference \cite{Nery2022} utilizes an electron-phonon coupling parameter $\eta$, which in our model corresponds to $\eta = bt_0/d_0$. We reproduced their Figure 8 results using the same parameter set of $\Delta_{\alpha}=0$, $t_0=2.58eV$ and $d_0=1.413$\AA\  for two different coupling strengths, $\eta_{\text{expt}} = 4.42$ eV/\AA (a value extracted from experiment), and $\eta_{\text{inter}}=2.5\eta_{\text{expt}}$ by running MTEF dynamics on a $36\times 36$ $\mathbf{k}$ and $\mathbf{q}$ grid for $12$fs. We calculate the retarded Green's function for band $n$ via:
\begin{equation}
    G_{W,n\mathbf{k}}(t) = \frac{i}{N_t}\sum_{il}\braket{n\mathbf{k}|\psi_l^i(t)}\braket{\psi_l^i(t=0)|n\mathbf{k}}.
\end{equation}
By taking the Fourier transform of the mean field propagated Green's function, using the mask function $W(x) = 1-3x^2 + 2x^3$, we obtain $G_{W,n\mathbf{k}}(\omega)$. The spectral function is of course defined to be $A_{n\mathbf{k}}(\omega) = -(1/\pi)\text{Im}\left[G_{W,n\mathbf{k}}(\omega)\right]$. The results shown in Fig. \ref{fig:Mauri}, have good agreement with the Nery and Mauri results, namely a significant broadening of the signal with increasing temperature and coupling strength, as well as a decrease in signal intensity away from K and $\Gamma$ for both cases. The small negativities in the spectral function for high energy $\mathbf{k}$ points near $\Gamma$ in the lower panel are due to the extremely anharmonic dynamics arising from the artificially strong electron-phonon coupling constant. Such spectral negativity is a known feature of MTEF calculations with anharmonic forces \cite{Lively2021}. 
\begin{figure}
    \centering
    \includegraphics[width=0.8\textwidth]{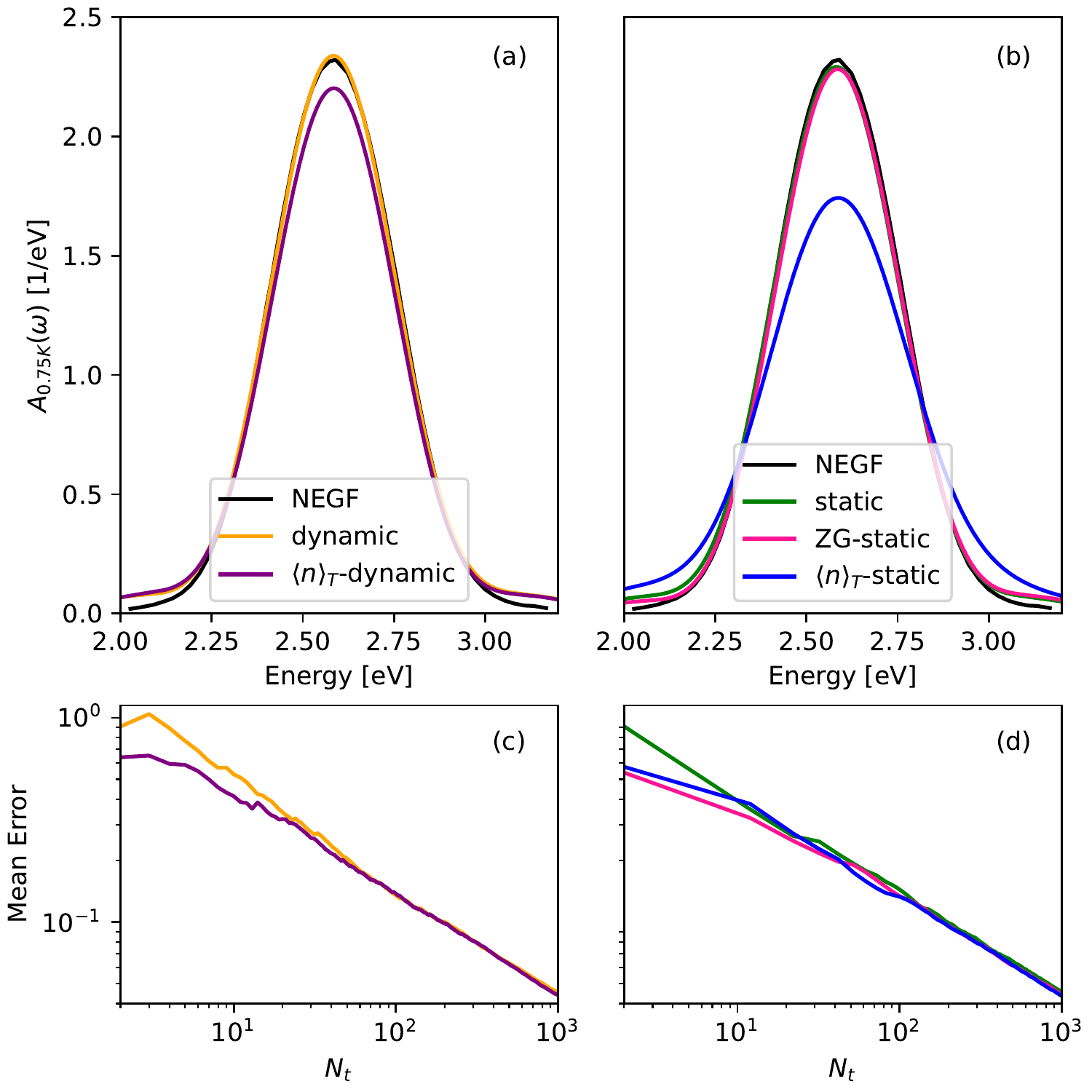}
    \caption{The spectral function of graphene for $\mathbf{k}=0.75K$, using $\eta_{\text{expt}}$ and sampling at $T=300K$ using dynamic phonon approaches in panel (a) and static phonon approaches in panel (b), with data taken from \cite{Nery2022}, Fig. 7, top left panel. The convergence of the spectral function for different sampling types in \ref{table:sampling types} is shown in panels (c) and (d) as the mean of the error of the time signal $G_{0.75K}(t)$ as a function of $N_t$, where $\text{err}(t)=\sigma_{G_{0.75K}}(t)/\sqrt{N_t}$ for $\sigma_G^2(t) = \braket{G^2(t)} - \braket{G(t)}^2$. }
    \label{fig:SI graphene 0.75K}
\end{figure}

We further compare our model to a specific peak at $\mathbf{k}=0.75K$ in Fig. \ref{fig:SI graphene 0.75K}. The various sampling approaches that are compared are explained in section \ref{section: SI Alternative Sampling Approaches}. In panels (a) and (b) we see that the sampling approaches used in the text, dynamic and static, agree very well with data extracted from \cite{Nery2022}, with the primary difference being slightly broader tails. With this check we are confident that simply replacing the parameters of our tight binding model with those fit to hBN is a reasonable approach to the electron-phonon coupling in this system. 

\section{Convergence of Valley Homogenization}\label{section: SI Convergence}
For this analysis we use the Normalized Root Mean Squared Displacement (NRMSD) to quantify how much a given time dependent signal varies from another:
\begin{equation}
    \text{NRMSD} = \frac{\sqrt{\int_{t_i}^{t_f} dt\left(f(t)-g(t)\right)^2/(t_f-t_i)}}{\text{max}\left(f(t)\right) - \text{min}\left(f(t)\right)}.
\end{equation}
We take $f(t)=\text{VA}_{N_t}(t)$ calculated with the largest number of trajectories available, in the case of the data in Fig. \ref{fig:valley-polarization-from-occupation}, $N_t=380$. We let $g(t) = \text{VA}_{N_t'}(t)$ be the same signal calculated with a random selection of $N_t'<<N_t$ trajectories. The NRMSD value for this collection of $N_t'$ trajectories tells us how much this signal deviates from the more converged value calculated with $N_t$ trajectories. By pulling different random collections of size $N_t'$ from the data we can histogram the resulting NRMSD values, giving us a probability distribution of what errors we can expect with this number of trajectories. The results of this analysis are plotted in Fig. \ref{fig: depol convergence}. 

Plotted in panel (a) in black is the same MTEF dynamic phonon data plotted in Fig. \ref{fig:valley-polarization-from-occupation}(b), alongside an example of a signal with an NRMSD of 0.02 in red. It's clear that this signal already appears qualitatively converged, yet we see from the error probability distributions in panel (b) that this a high error outlier of a result, even when using only two trajectories! Instead, with quite reasonable numbers of samples, the results rapidly converge towards signals which are graphically indistinguishable from results using over an order of magnitude more trajectories. 
\begin{figure}
    \centering
    \includegraphics[width=\linewidth]{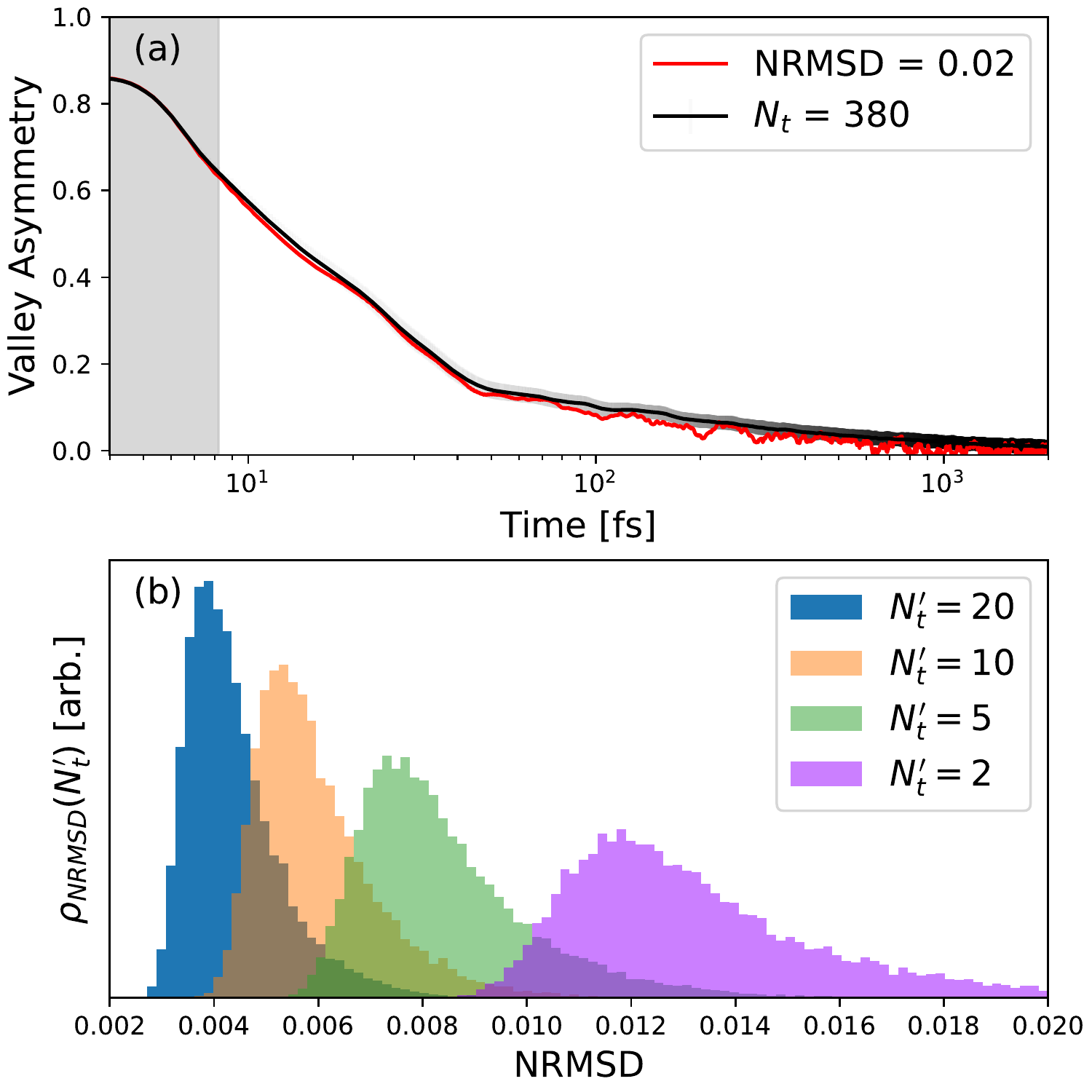}
    \caption{The probability distribution of getting a particular error measured as the NRMSD of the valley polarization seen in Fig. \ref{fig:valley-polarization-from-occupation} for a small number of samples with respect to the value calculated using a much larger number of samples.} 
    \label{fig: depol convergence}
\end{figure}

\section{Alternative Sampling Approaches}\label{section: SI Alternative Sampling Approaches}
In the special displacement method (SDM) of Zacharias and Giustino \cite{Zacharias2016,Zacharias2020}, the convergence of sampling the configuration distribution in Eq. (\ref{eq:ZG average}) can be accelerated for small supercell sizes (i.e. sparse $\mathbf{q}$ grids in the BZ) by taking a handful of specific configurations. These involve taking positions of magnitude $|x_{\mathbf{q}\nu}|,|y_{\mathbf{q}\nu}| =\sigma_{\mathbf{q}\nu}$ and carefully choosing signs $S_{\mathbf{q}\nu}=\{+1,-1\}$ of $z_{\mathbf{q}\nu} = S_{\mathbf{q}\nu} |z_{\mathbf{q}\nu}|$, which favorably cancel terms in the 2nd order perturbative expansion of observables in the phonon coordinates. This allows for capturing the thermodynamic equilibrium properties at second order which can be shown to analytically go to zero in the case of an infinitely large supercell. This is expected to be of most benefit in the case of small supercells, and has been found to not to necessarily be a sufficient configuration for convergence in cases of very strong electron-phonon coupling \cite{Nery2022}.

In our framework we can analyze this choice of displacement magnitude in terms of the phonon occupation number. Taking equations (\ref{eq:phonon hamiltonian ladder operators}), we clearly identify $n_{\mathbf{q}\nu}$ to be
\begin{equation}\label{eq:SI phonon number}
\begin{split}
n_{\mathbf{q}\nu}^x &= 
\frac{1}{2}\left(\tilde{r}_{\mathbf{q}\nu}^2 + \tilde{x}_{\mathbf{q}\nu}^2\right) - \frac{1}{2}\null\quad\text{for } \mathbf{q}\in \mathcal{A}, \mathcal{B}\\
n_{\mathbf{q}\nu}^y &=\frac{1}{2}\left(\tilde{s}_{\mathbf{q}\nu}^2 + \tilde{y}_{\mathbf{q}\nu}^2\right) - \frac{1}{2}\null\quad\text{for } \mathbf{q}\in \mathcal{B}.\\
\end{split}
\end{equation}
In this picture, with the SDM choice equivalent to setting the reduced momenta $\tilde{r}, \tilde{s}$ to zero and $\tilde{x}^2 = \tilde{y}^2=n_{\mathbf{q}\nu,T}+1/2$ meaning $n_{\mathbf{q}\nu} = n_{\mathbf{q}\nu,T}/2 - 1/4$. That is to say, the SDM corresponds to a static displacement large enough to account for half of the thermal occupation of the phonon mode, and half of the ZPE. 

We can perform an analogous sampling which obtains the thermal occupation number by construction but splits the weight evenly between position and momentum by choosing $\tilde{r}_{\mathbf{q}\nu}^2=\tilde{x}_{\mathbf{q}\nu}^2=n_{\mathbf{q}\nu,T}^x + 1/2$, and equivalently for $n^y_{\mathbf{q}\nu}$. Since in our approach, we are interested in strongly coupled electron-phonon systems where a perturbative expansion may not be sufficient, and we have dynamical phonons which can explore beyond a harmonic approximation, rather than carefully choosing signs to eliminate terms for a 2nd order thermodynamic equilibirum property, we simply randomly sample $S_{\mathbf{q}\nu}$ for every phonon branch and momenta. We refer to this as $\langle n\rangle_T$ sampling. 

If on the other hand, we choose to run with frozen phonon dynamics, but still want to have initial phonon occupations corresponding to the thermal occupation, we can set $\tilde{x}_{\mathbf{q}\nu}^2=2n_{\mathbf{q}\nu}^x + 1$ and sample the sign of $\tilde{x}_{\mathbf{q}\nu}$ (doing the same with $\tilde{y}$), while fixing $\dot{\tilde{z}}_{\mathbf{q}\nu}=0$. We refer to this approach as $\langle n\rangle_{T}$-static. For comparison, we refer to using the SDM magnitudes ($n_{\mathbf{q}\nu}=n_{\mathbf{q}\nu,T}/2$) and sampling signs with frozen phonons as ZG-static. A summary of the various initial conditions and dynamics choices is shown in Table \ref{table:sampling types}.
\begin{table}\label{table:sampling types}
\begin{tabular}{|l|l|l|l|}
\hline
                            & $\tilde{x}_{\mathbf{q}\nu},\ \tilde{y}_{\mathbf{q}\nu}$ & $\tilde{r}_{\mathbf{q}\nu},\ \tilde{s}_{\mathbf{q}\nu}$ & $n_{\mathbf{q}\nu}$    \\ \hline
dynamic                       & $\sim \rho_{\text{ph}, W}$                                  & $\sim \rho_{\text{ph}, W}$                                  & $\to n_{\mathbf{q}\nu,T}$\\ \hline
$\langle n\rangle_T$-dynamic        & $\pm(n_{\mathbf{q}\nu} + \frac{1}{2})^{1/2}$                           & $\pm(n_{\mathbf{q}\nu} + \frac{1}{2})^{1/2}$                           & $n_{\mathbf{q}\nu,T}$   \\ \hline
static                       & $\sim \rho_{\text{ph}, W}$                           &  0                                                           &  $\to n_{\mathbf{q}\nu,T}/2-1/4$   \\ \hline
ZG-static                   & $\pm(n_{\mathbf{q}\nu} + \frac{1}{2})^{1/2}$                           & 0                                                           & $n_{\mathbf{q}\nu,T}/2-1/4$ \\ \hline
$\langle n\rangle_T$-static & $\pm(2n_{\mathbf{q}\nu} + 1)^{1/2}$                                    & 0                                                            & $n_{\mathbf{q}\nu,T}$  \\ \hline
\end{tabular}
\caption{A summary of the various choices of initial conditions for the phonon reduced coordinates and momenta. Although the electronic system properties can be obtained via time evolution for all these methods, the phonon system is dynamic only when $\tilde{r}_{\mathbf{q}\nu},\tilde{s}_{\mathbf{q}\nu}\neq 0$. The symbol ``$\to n_{\mathbf{q}\nu,T}$" indicates convergence with increasing $N_t$.
}
\end{table}
\begin{figure}
    \centering
    \includegraphics[width=0.8\textwidth]{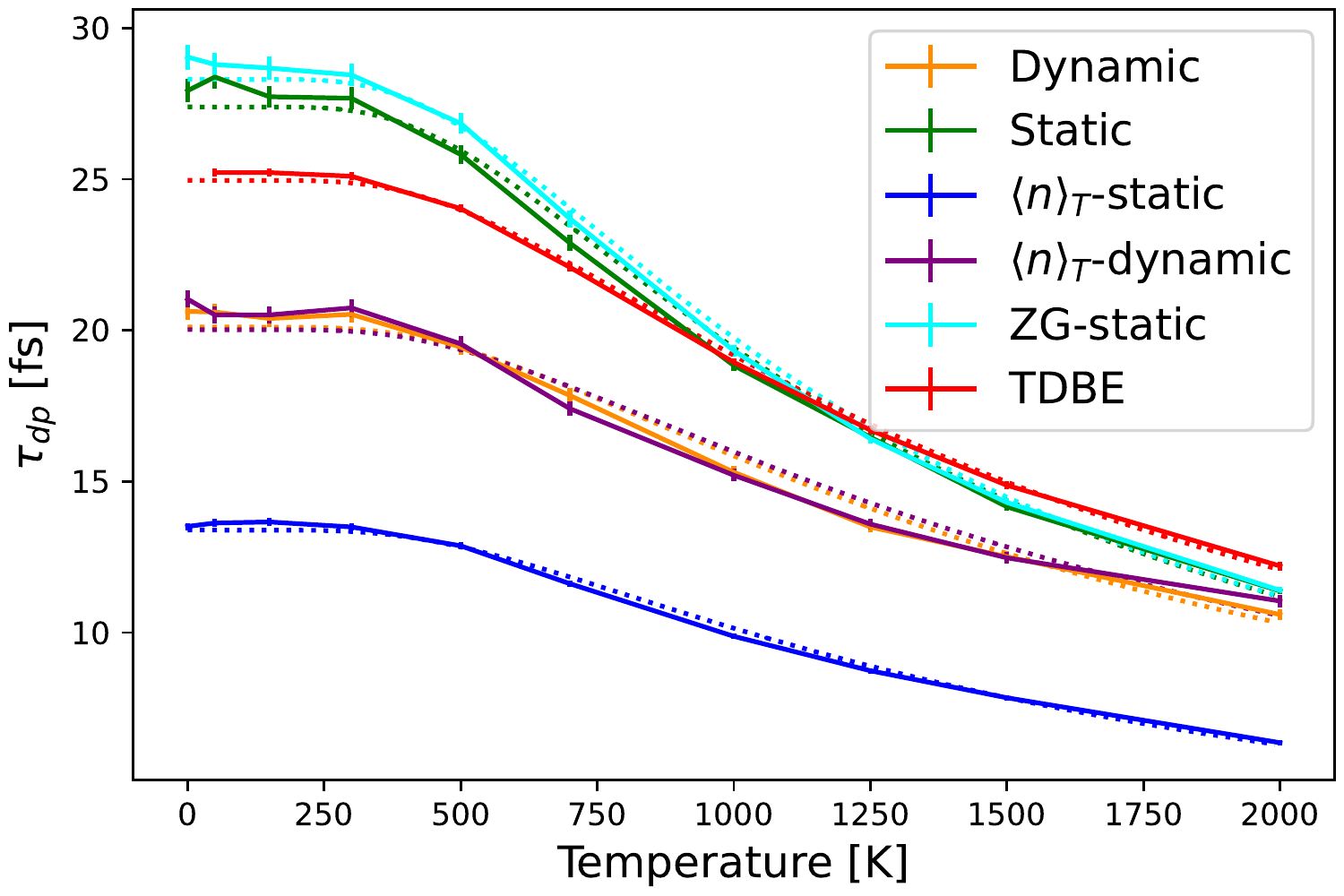}
    \caption{ The temperature dependence of the short time scale valley population homogenization, as seen in Fig. \ref{fig:valley-polarization-from-occupation}, calculated with all of the sampling techniques in Table \ref{table:sampling types}. 
    }
    \label{fig:SI valley-polarization-from-occupation}
\end{figure}

The principle question is whether these different sampling techniques lead to altered observables, or faster convergence. We start by looking at a phonon renormalized property of the electronic system in a thermal equilibrium state. Taking the graphene spectral function in Fig. \ref{fig:SI graphene 0.75K}(a), we see that using `dynamic' sampling, that is drawing phonon momenta and coordinates in a straight-forward manner from Eq. \ref{eq:phonon distribution} and $\langle n\rangle_T$-dynamic already lead to slightly different spectral function intensities. In Fig. \ref{fig:SI graphene 0.75K}(c), we track the convergence of this signal by taking the average error as a function of time. Given that we are Monte Carlo sampling an initial distribution this error is defined as the standard deviation of the observable calculated with $N_t$ trajectories, over $\sqrt{N_t}$. Panel (c) shows that the signal produced by $\langle n\rangle_T$ sampling converges marginally faster than that produced by dynamic sampling, but that at large configurations, there is virtually no difference in the two approaches. 

We see something similar when comparing the `static' and `ZG-static' sampling approaches in Fig. \ref{fig:SI graphene 0.75K}(b) and Fig. \ref{fig:SI graphene 0.75K}(d), with the primary difference of this comparison being that the spectral function magnitudes for these two methods agree nearly exactly. The outlier however appears to be $\langle n\rangle_T$-static, which has a significantly different spectral intensity than any static or dynamic approach. This indicates that the majority of the effect responsible for renormalization of the electronic spectral function at this $\mathbf{k}$ point can be accounted for via displacement within the first standard deviation of the phonon distribution. Of course in the harmonic limit, setting phonon momentum and displacement absolute values to $\pm|\sigma_{\mathbf{q}\nu}|$ will result in motion in phase space entirely on the circumference of the phase space circle of radius $\sqrt{2}|\sigma_{\mathbf{q}\nu}|$. Fixing the initial position to $\pm2|\sigma_{\mathbf{q}\nu}|$ apparently exceeds the renormalization effect obtained by full Monte Carlo sampling, or even Monte Carlo sampling on the $\pm\sqrt{2}|\sigma_{\mathbf{q}\nu}|$ radial points.

However, this is a value calculated at thermodynamic equilibrium. To see the effect for far from equilibrium properties, we turn to the temperature dependence of the valley homogenization timescale in the TB hBN model, seen in Fig. \ref{fig:SI valley-polarization-from-occupation}. The $\langle n\rangle_T$-dynamic and ZG-static sampling approaches sit almost precisely on top of their coresponding dynamic and static results. Quite notably the $\langle n\rangle_T$-static results again constitute an outlier to the other methods, with a scattering fit parameter of $\alpha_{\langle n\rangle_T-\text{static}}=0.129$fs$^{-1}$, compared to  $\alpha_{\text{Static}} = 0.081 \text{fs}^{-1}$, $\alpha_{ZG-\text{static}}=0.082$fs$^{-1}$, $\alpha_{\text{Dynamic}} = 0.072 \text{fs}^{-1}$, $\alpha_{\langle n\rangle_T-\text{Dynamic}} = 0.068 \text{fs}^{-1}$ and $\alpha_{\text{TDBE}} = 0.065 \text{fs}^{-1}$. Clearly this sampling method, while producing the correct thermal occupation value through strong static displacement, appears to overestimate the effect of phonon renormalization, when compared to Monte Carlo sampling of the phonon coordinate / momenta distribution. 

\begin{figure}
    \centering
    \includegraphics[width=0.8\textwidth]{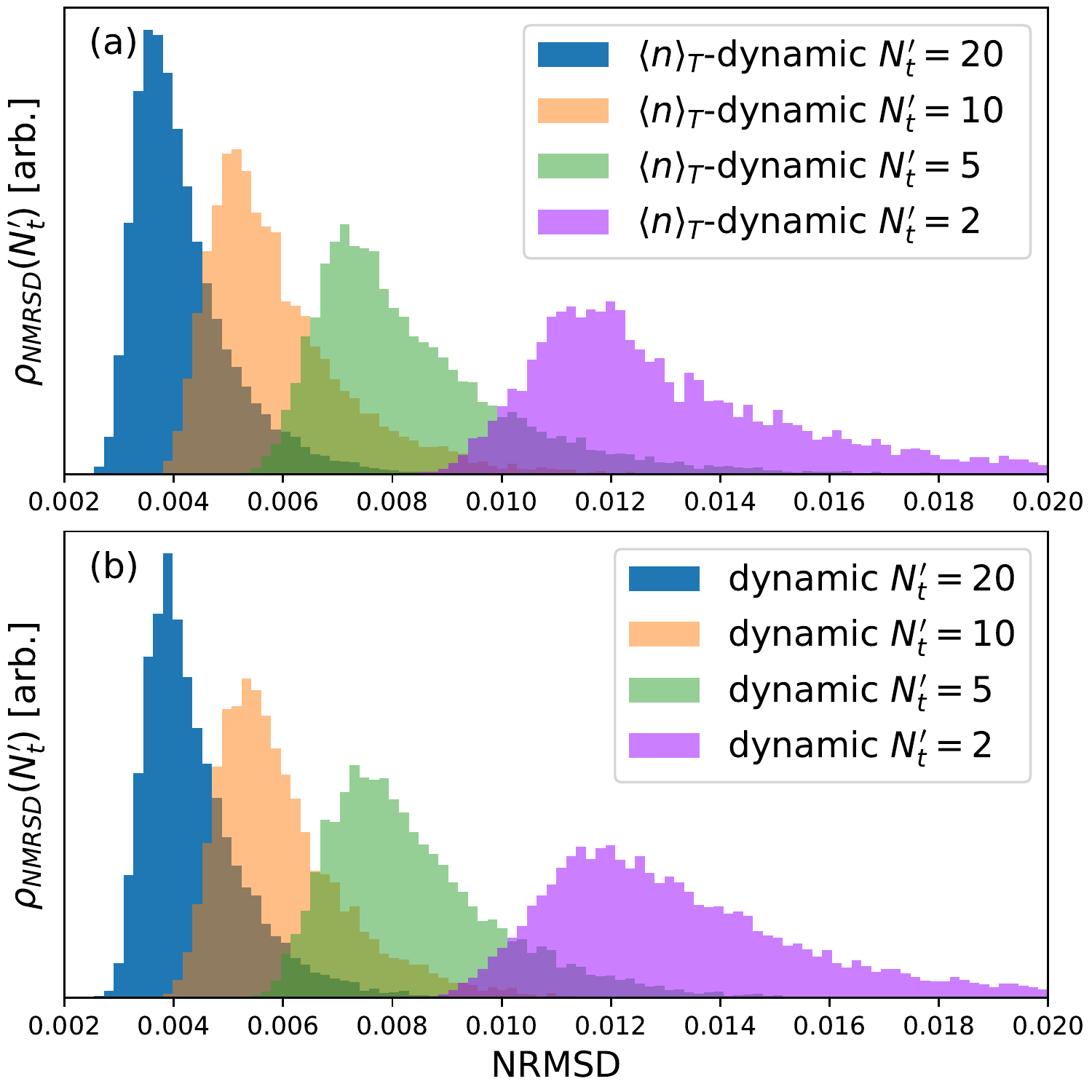}
    \caption{The probability distribution of getting a particular NRMSD of the valley polarization seen in Fig. \ref{fig:valley-polarization-from-occupation} when selecting an arbitrary $N_t$ number samples taken via $\langle n\rangle_T$-dynamic sampling versus straight forward Monte Carlo sampling.} 
    \label{fig:SI Valley asymmetry convergence}
\end{figure}
Finally we can investigate whether there is any significant advantage in the dynamical case to sampling with the $\langle n\rangle_T$-dynamic approach. We repeat the NRMSD probability analysis of section \ref{section: SI Convergence} for the time dependent valley asymmetry in an MTEF calculation and report the results in Fig. \ref{fig:SI Valley asymmetry convergence}. We find that the probability of obtaining a signal using a small number of trajectories, which has a small deviation from the signal one would obtain using a much larger number, looks effectively identical between the two sampling approaches, thus in this case this sampling method confers no appreciable advantage. 

\section{ZPE Leakage}\label{section: SI ZPE Leakage}
We track the effects of ZPE leakage when propagating with dynamic MTEF by looking at the phonon occupation number $n_{\mathbf{q}\nu}$ calculated via Eq. (\ref{eq:SI phonon number}). Given sufficient samples, $n_{\mathbf{q}\nu}$ will converge towards the expected thermal occupation $n_{\mathbf{q}\nu,T}$, which for the optical phonons at 300K means approximately 0. Therefore as the ZPE drains out of a given phonon mode, the occupation will go towards $-1/2$.

In Fig. \ref{fig:SI ZPE loss} we show a selection of the occupation numbers of the highest energy optical branch phonon modes over time. In panel (a) when the system is exposed to a laser pulse, we see that on timescales commensurate with the long time scale excitation outside the valleys in Fig. \ref{fig:valley-polarization-from-occupation}(b), there is a loss of energy in the phonon modes. The initial occupation value of 0 requires a large number of trajectories to converge to exactly, but as discussed throughout the text, the dynamical electronic observables of interest converge rapidly. Panel (b) shows the phonon occupation when the electronic system is not pumped. In this case, due to the large electronic gap, there is nowhere for the phonon energy to go, and instead one just sees oscillation of the occupation numbers over time. This oscillation is due to anharmonic forces arising from exposure to the electronic system, and can in principle be analyzed to capture the renormalized phonon frequencies. 

In Fig. \ref{fig:SI ZPE loss}(c) we take the phonon occupations from the last time step of panel (a) at $2$ps and plot them against the distance of their $\mathbf{q}$ vector from $\Gamma$. There is very clearly a direct correlation between the how close a phonon mode is to $\Gamma$ and the amount of energy it loses to the electronic system. This may be related to the fact that the phonon modes are coupled directly to the electronic system via a linear dependence through the nearest neighbor hopping term. Therefore the modes closer to $\Gamma$ which correspond to a coherent reduction in the nearest neighbor distance throughout the system, most strongly excite the electrons and allow a conduit for vibrational energy to go into the electronic system. 
\begin{figure}
    \centering
    \includegraphics[width=\textwidth]{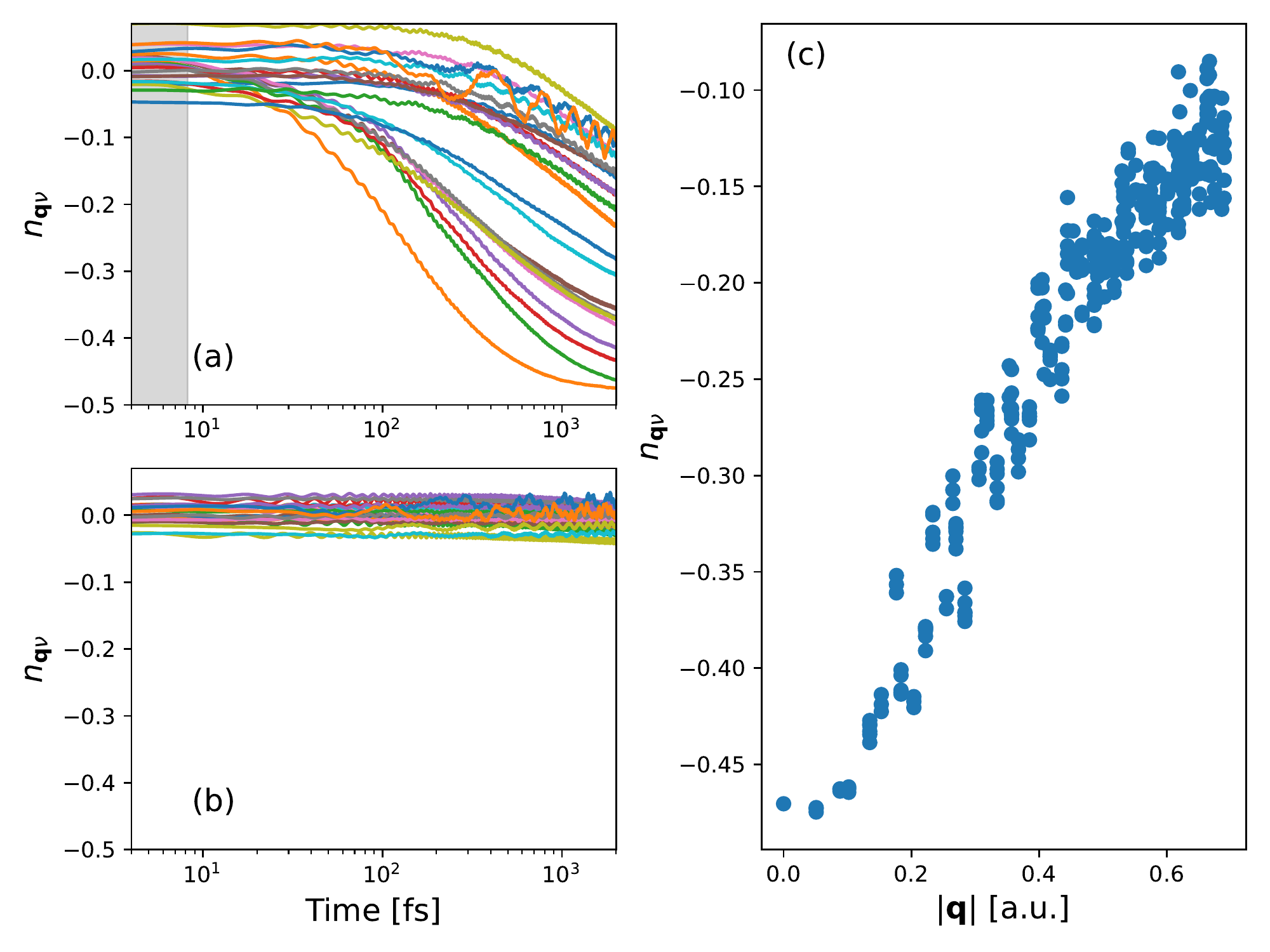}
    \caption{The phonon occupation number in a selection of the highest energy optical modes over time, when the electronic system is either (a) exposed to a pump or (b) allowed to propagate without a pump. Panel (c) shows the phonon occupation numbers from panel (a) at the final time, plotted against the distance of their $\mathbf{q}$ vector from $\Gamma$. }
    \label{fig:SI ZPE loss}
\end{figure}

\end{widetext}


\end{document}